\documentclass[twocolumn]{aastex631}
\let\splitbox\relax
\usepackage{makecell,lmodern}
\usepackage{csquotes}
\usepackage{circuitikz}

\usepackage[T1]{fontenc}

\newcommand{\ice}{H$_2$O}
\newcommand{\tauice}{$\tau_{\rm ice}$}

\newcommand{\DEL}[1]{}

\usepackage{verbatim}
\usepackage[T1]{fontenc}
\usepackage{braket}
\usepackage{adjustbox}
\usepackage{xpatch}

\usepackage{hyperref}
\usepackage{float}
\usepackage{graphicx}	
\usepackage{amsmath}	
\usepackage{booktabs,multirow}
\usepackage{soul}
\usepackage[table]{xcolor}
\usepackage{changepage,threeparttable}
\usepackage[shortlabels]{enumitem}

\usepackage{amsmath}
\usepackage{graphicx}
\usepackage{braket}
\usepackage{xcolor}

\begin{document}

   \title{Using Scattered Near-Infrared Light to Map Water Ice in Prestellar Cores with SPHEREx}

    \author[0009-0003-9906-2745]{Tamojeet Roychowdhury}
    \affiliation{Department of Astronomy, University of California Berkeley, Berkeley, CA 94704, USA}
    \email{tamojeet@berkeley.edu}
    
    \author[0000-0002-8716-0482]{Jennifer B. Bergner}
    \affiliation{Department of Chemistry, University of California Berkeley, Berkeley, CA 94704, USA}

    \author{Jens Kauffmann}
    \affiliation{Haystack Observatory, Massachusetts Institute of Technology, 99 Millstone Road, Westford, MA~01886, USA}

    \author{Thushara G.S. Pillai}
    \affiliation{Haystack Observatory, Massachusetts Institute of Technology, 99 Millstone Road, Westford, MA~01886, USA}

    \author{Silvia Spezzano}
    \affiliation{Max-Planck-Institut für Extraterrestrische Physik, Giessenbachstrasse 1, 85748 Garching, Germany}


   \begin{abstract}
   We present the first coreshine-derived, spatially-resolved maps of the 3 $\mu$m H$_2$O ice absorption feature in four prestellar cores, using SPHEREx spectra. Ices are a key component of dense cores in molecular clouds, playing a central role in the chemistry of planet formation around young stars. However, spatially resolved abundance studies remain limited, typically relying on unevenly distributed background star sightlines. Here, we take advantage of the all-sky spectrophotometric capabilities of SPHEREx to construct ice absorption maps with uniform spatial resolution using the illumination of dense cores by scattered Galactic radiation, or coreshine. To demonstrate proof of concept, we analyse the spatially varying H$_2$O ice absorption in four nearby ($\sim140$ pc) prestellar cores -- L1544, CrA 151, L260 and L1512. Two cores follow the expected spatial trend of ice absorption peaking at the centre, but the two densest cores show a surprising drop in observed ice absorption in the innermost regions. To interpret the absorption maps, we construct analytical and simulated models of a Bonnor-Ebert sphere illuminated by scattering. We study the effects of different geometric configurations, ice mass fractions, and spatial differences in ice composition. None of these can explain the reduced central absorption, pointing to an unexplained physical or chemical effect operating in the densest prestellar regions. Our simulations further show that spectra derived from coreshine provide a robust tracer of spatially varying ice density and composition, establishing SPHEREx scattered-light spectroscopy as a powerful new probe of ice in dense cores. 
\end{abstract}


   
%
\section{Introduction}
H$_2$O ice is one of the most important tracers of chemistry and physics in star- and planet-forming regions \citep[for reviews see][]{Boogert2008, Oberg2011, Boogert2015, Dishoeck2021, Cuppen2024}, and is widely detected via its prominent absorption band centred around 3 $\mu$m \citep{Leger1979}. Its abundance is also an important reference against which ice-phase abundances of other molecular species, like CO, CO$_2$, NH$_3$ and CH$_3$OH are measured \citep[for example][]{Pontoppidan2008, Oberg2008, Bottinelli2010, Boogert2015, Oberg2021, Brunken2024, Dishoeck2025}. During planet formation, molecular ices also act as reservoirs for volatile materials and prebiotic building blocks \citep{Dishoeck2014, Boogert2015, Oberg2021, Merel2026}.

Water ice has been observed in a multitude of star-forming environments via its 3 $\mu$m absorption feature, thanks in large part to prior space-based missions such as the Infrared Space Observatory \citep[see][for a review]{Gibb2004} and \textit{Spitzer} \citep[see][for a review]{Boogert2015}. From this, it is known that water ice forms beyond an extinction threshold of $A_V \approx 1.6$, and remains largely frozen throughout the dense stages of star formation, increasing the total mass of solids and sequestering the majority of the cosmic oxygen budget in grains \citep{Hollenbach2009, Boogert2015}. 

However, to date, virtually all of our knowledge of ice formation and evolution in prestellar cores relies on line-of-sight absorption toward background stars \citep{Whittet1988, Murakawa2000, Knez2005, Bergin2005, Pontoppidan2004, Pontoppidan2006}. These studies initially achieved poor spatial resolution ($\sim 0.1^\circ$) or relied on $\lesssim 10$ sightlines in smaller-scale environments. These two limitations severely hinder our understanding of how the local physical conditions influence the formation and evolution of ices. More recently, stellar sightline-based absorption maps have achieved higher resolution and probe higher column densities, often for multiple ice species \citep{Noble2013, Noble2017, Perotti2020, Goto2021}, while  JWST has further improved the spatial resolution and enabled several tens of sightlines to be detected on core scales \citep{Smith2025}.  Still, the smaller field-of-view and extended integration times needed to probe the highly extincted background stars makes large-scale spectroscopic mapping prohibitively resource-intensive.

The launch of the Spectro-Photometer for the History of the Universe, Epoch of Reionization, and Ices Explorer mission (SPHEREx, \citealt{Bock2025}) in 2025 has provided a new opportunity to map ice compositions in prestellar environments across unprecedentedly large spatial scales. SPHEREx is designed to image the entire sky in 102 narrow photometric bands in the 0.75--5 $\mu$m range, effectively functioning as a low-resolution, all-sky spectrograph. This opens up an immediate avenue to use SPHEREx spectra of background stars to map out molecular cloud and protostellar chemical abundances \citep{SPLICES2023, Melnick2026} on large scales and reaching resolutions limited only by the separation between stars behind the cloud. However, this approach is still constrained to only specific sightlines with background stars. Moving beyond this, \cite{Hora2026} recently used SPHEREx spectra to construct large-scale ice abundance maps across a $\sim 5^\circ\times 5^\circ$ region of the Cygnus-X region using absorption features observed against the diffuse Galactic background -- similar to a \textit{Spitzer} study of Cepheus A by \cite{Sonnentrucker_2008}.

On smaller scales ($0.1-1$ pc), a significant fraction of dense star-forming cores are also known to have micron-sized dust grains that can scatter the ambient Galactic radiation field and illuminate the core (coreshine, \citealt{Pagani2010, Steinacker2015, Andersen2013}). In this work, we present the first set of water ice absorption maps viewed against the diffuse coreshine, in four well-studied nearby cores. To our knowledge, this is the first study to derive spatial variations in the 3 $\mu$m H$_2$O ice band purely from diffuse scattering spectra across multiple dense cores, instead of discrete stellar sightlines. This approach will open the door to answering key questions on the interplay of physical and chemical properties in different regions of molecular clouds and in different phases in the timeline of core collapse and protostar formation. Here we primarily focus on identifying how the ice abundance varies with gas density from the inner core to the outer regions, and whether all cores follow the same spatial variation patterns. 

In \S~\ref{sec:observations} we introduce the sample and the method to create spatial maps of ice absorption from SPHEREx data. In \S~\ref{sec:model} we first introduce a basic analytical model for near-infrared (NIR) scattering in a prestellar core to help interpret the observed water ice variations. We then run full 3D Monte Carlo radiative transfer calculations to demonstrate that scattered NIR flux indeed probes ice absorption within the cloud. In \S~\ref{sec:discussion} we study the effects of varying geometry, dust grain growth and changes in ice composition at high densities on the observed water ice absorption map. We also attempt to reproduce the spatial trends in observed ice maps and compare to what is known from studies on similar physical scales in the literature. In \S~\ref{sec:conclusions} we conclude with a brief discussion and summary of our findings.

\section{Observational Data and Results}
\label{sec:observations}
\subsection{Target Selection}
\label{subsec:targets}
We base our analysis on four relatively nearby and well-studied cores outlined in Table~\ref{tab:core-sample}. Three of these were selected from the core samples showing prominent coreshine in \textit{Spitzer} 3.6 $\mu$m images, from \cite{Pagani2010} and \cite{Steinacker2015}, focussing only on the ones that are close ($<200$ pc in distance), have a roughly spherical or simple profile (i.e. avoiding highly elliptical or double-lobed cores), do not have an immediately neighbouring bright protostar considerably influencing coreshine, and are visually prominent in the NIR in SPHEREx images. 

Two of the studied objects, L1512 and L260, have typical densities expected of starless cores ($\rho_{\rm central} \sim 10^4-10^5$ cm$^{-3}$, \citealt{Steinacker2015, Jensen2024}), while the third, L1544, has one of the highest-densities known ($\rho_{\rm central} \sim 10^6-10^7$ cm$^{-3}$ in the innermost 500--1000 AU, \citealt{Keto2014}). We added CrA 151 as a second object in the high density set ($\rho_{\rm central} \sim 10^7$ cm$^{-3}$ in a similar radius, \citealt{Redaelli2025}) to facilitate comparison, which may potentially host a protostar but is primarily illuminated by coreshine. Our sample is not meant to be exhaustive, but since the objective of this paper is to demonstrate its utility to probe ices and find spatial trends, the exact sample is not important for the proof-of-concept presented here.  

\begin{table*}
\begin{center}
\begin{tabular}{ccccccc} \hline
Name & RA & Dec & Star-forming region &  Central Density & \tauice\ at 3 $\mu$m \\ 
& [deg] &  [deg]  & & $\rho_0$ [cm$^{-3}$] & ($\tau\approx\kappa\rho_0r_0$)$^{(a)}$ \\ \hline
L1512 & 76.0364167 & 32.7211667 & Auriga & $10^4-10^5$ & 0.03--2.7 \\ 
L260 & 251.7885000 & -9.5835556 & Ophiuchus & $10^4-10^5$ & 0.03--2.7 \\ 
L1544 & 76.0697917 & 25.18125 & Taurus & $10^6-10^7$ & 2.7--270 \\ 
CrA 151 & 287.5857417 & -37.1408333 & Corona Australis &  $10^7$ & 27--270 \\ 
\hline
\end{tabular}
\end{center}
\caption{Sample of cores studied in this work. Density references are given in \S~\ref{subsec:targets}. (a) For each core we also compute the corresponding total line-of-sight H$_2$O ice opacity at 3 $\mu$m, \tauice\, through the central region. This column gives a range of \tauice\ values assuming a Bonnor-Ebert radius $r_0$ of 0.002--0.02 pc (described in detail in \S~\ref{sec:model}) and dust opacity $\kappa$ derived from \texttt{OpTool} (also detailed in \S~\ref{sec:model}).}
\label{tab:core-sample}
\end{table*}

\subsection{Ice Depth Estimation using SPHEREx Photometry}

SPHEREx provides photometry in 102 bands in the wavelength range of 0.75--5 $\mu$m \citep{Bock2025}. We use their Python-based API via IRSA \citep{SPHEREx-IRSA} to retrieve and download all files that overlap within 1 arcsecond of the central position of each target in the detector 4 and 5 (D4 and D5) bands. These respectively cover $\sim2.4-3.8\ \mu$m and $\sim3.8-4.4\ \mu$m in wavelength and thus cover the \ice\ ice band along with its surrounding continuum. Each core had between 18 and 22 available images in each of D4 and D5. Zodiacal light (provided with the data files) is subtracted from each image before further processing.

SPHEREx uses linear variable filters (LVFs) placed directly in front of its detector arrays, such that each pixel captures light from a distinct sky position in a narrow photometric band whose central wavelength varies continuously across one axis of the focal plane. To create the spectrum for a particular coordinate, flux at different wavelengths from different images needs to be stitched together. We use the various FITS extensions in the data products to calibrate the wavelengths to each pixel \footnote{\url{https://caltech-ipac.github.io/irsa-tutorials/spherex-intro/}}. To reduce the effect of photometric uncertainties, images are smoothed using a 2D Gaussian kernel of $\sigma=1$ pixel before constructing spectra. 

We create a grid of area 7.2 arcmin $\times$ 7.2 arcmin centred at the coordinates of each core (in Table~\ref{tab:core-sample}) and with points spaced 7.2 arcseconds apart. This is slightly larger than the SPHEREx native pixel size of 6.2 arcseconds in order to avoid multiple grid points mapping to fluxes from the same SPHEREx pixel. For our cores, 7.2 arcseconds roughly corresponding to a physical size of 0.0052 parsecs or 1050 AU at a distance of 140 pc. For each grid point, we find the nearest pixel in each image and store the corresponding wavelength and flux as part of the spectrum. 

For some coordinates, the nearest-pixel-mapping technique results in a spectrum where most points trace the diffuse NIR flux but some points intersect the edge of the PSF of stellar sightlines (that dominates the pixel flux). These appear as sharp spikes in the spectrum and can strongly bias the continuum fitting in the latter steps. To mitigate this, in each spectrum, we remove any data point whose flux value is more than 1.5 times the median flux in the $2.4-2.7$ $\mu$m and $3.4-4.1$ $\mu$m ranges (with 1.5 being chosen by visually inspecting the retrieved spectra from multiple points around the targeted cores). We also remove data points that have flux measurements with signal-to-noise ratio $\leq 3$ (where the noise is computed from the variance map provided with the SPHEREx data). Note that this reduces the overall number of data points in a given pixel's spectrum, but does not render the pixel unusable for ice depth measurement. Similarly, lines of sight that are entirely pointed toward stars have spectra dominated by the stellar continuum (i.e. not coreshine) and can be used to quantify ice absorption as well. 

Since the spectral resolution of SPHEREx is relatively low, we need to ensure that the shape and depth of the ice band is adequately captured with our method. We first check that removal of S/N $\leq 3$ points does not remove the central region of the ice band (around 3 $\mu$m) for even the densest regions and regions with highest values of $\tau$ across our sample, which otherwise could cause the ice depth to be systematically underestimated. Also, \cite{Hora2026} compute correction factors in true depth vs apparent depth of various emission/absorption features due to low-resolution sampling of the line. For H$_2$O ice, they find the factor to be $\approx 1-1.1$; in other words, coarse wavelength sampling introduces at most a 10\% uncertainty into our calculated water ice optical depths.

We then estimate the continuum $F_{\mathrm{cont}}(\lambda)$ by fitting a quadratic function (typically also used for stellar sightlines, eg. \citealt{Oberg2011, Noble2013}) to the flux in wavelengths $2.4-2.7\ \mu$m and $3.6-4.1\ \mu$m. This avoids the full water ice absorption band and the prominent CO$_2$ band around $4.2\ \mu$m. It also largely avoids bands of methanol and hydrates of ammonia (see for example \citealt{JWSTref3}). The optical depth $\tau$ is calculated for each spectral point with respect to this continuum. The peak ice absorption depth, \tauice\ is estimated by taking all (usually two) flux points $F(\lambda)$ in a wavelength range of $2.9-3.1\ \mu$m and computing the average 

\begin{equation}\tau_{\rm ice} = \Braket{-\ln\frac{F(\lambda)}{F_{\mathrm{cont}}(\lambda)} } \label{eq:ice-depth}\end{equation} 

\begin{figure*}
 \centering
 \includegraphics[width=1.0\linewidth]{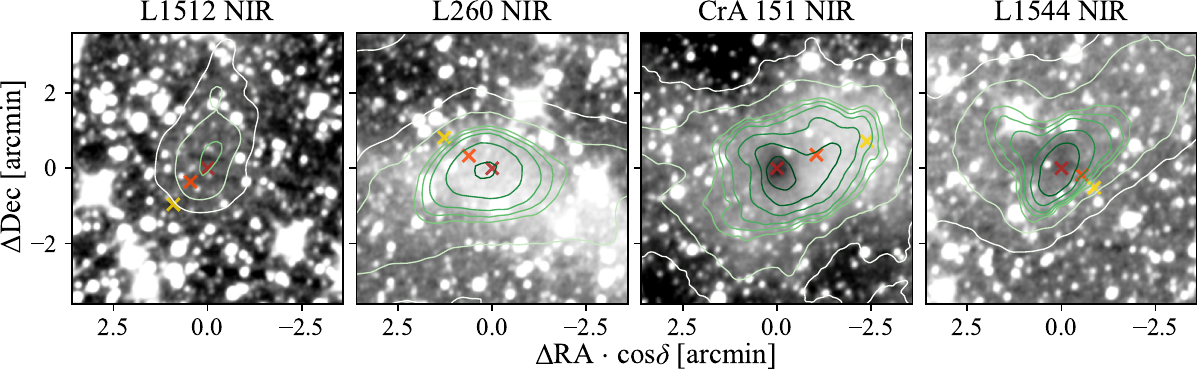}\vspace{15pt} \\ 
 \includegraphics[width=0.49\linewidth]{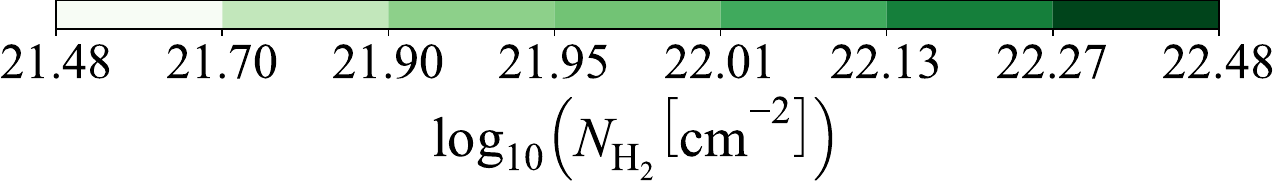}\ \ 
 \includegraphics[width=0.49\linewidth]{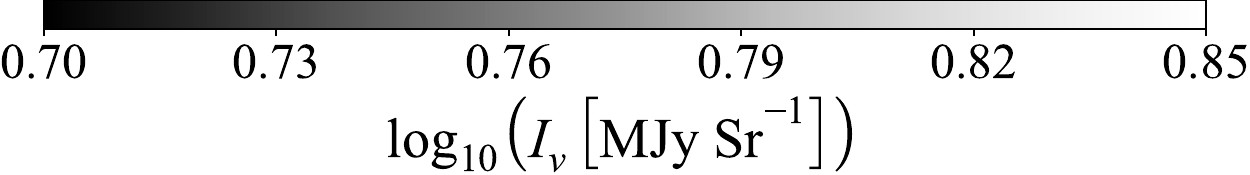}
 \vspace{15pt} \\
 \includegraphics[width=0.94\linewidth]{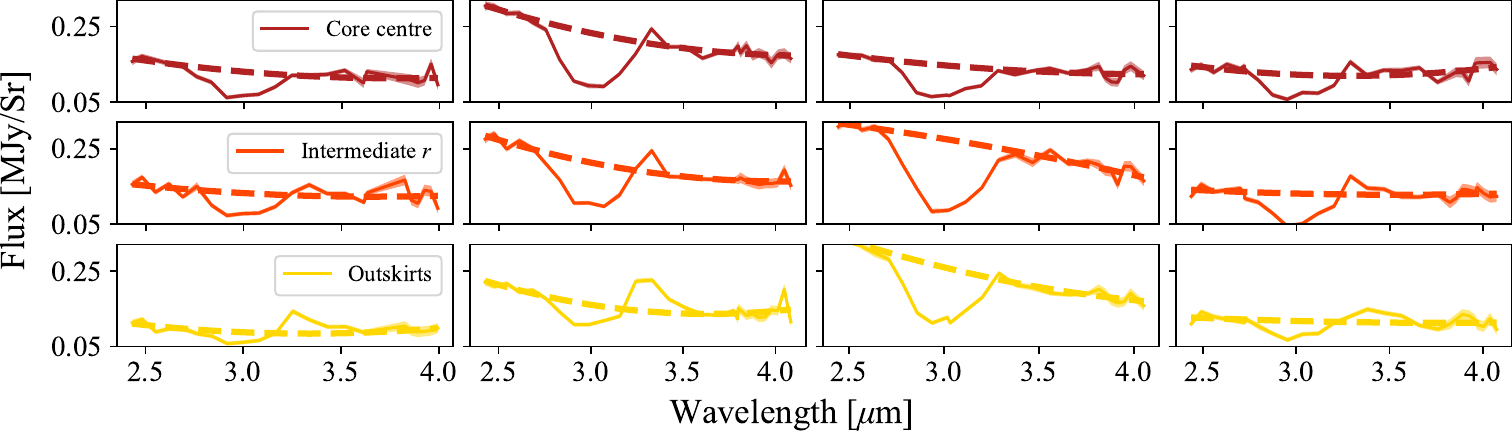}\vspace{15pt} \\ 
 \includegraphics[width=0.94\linewidth]{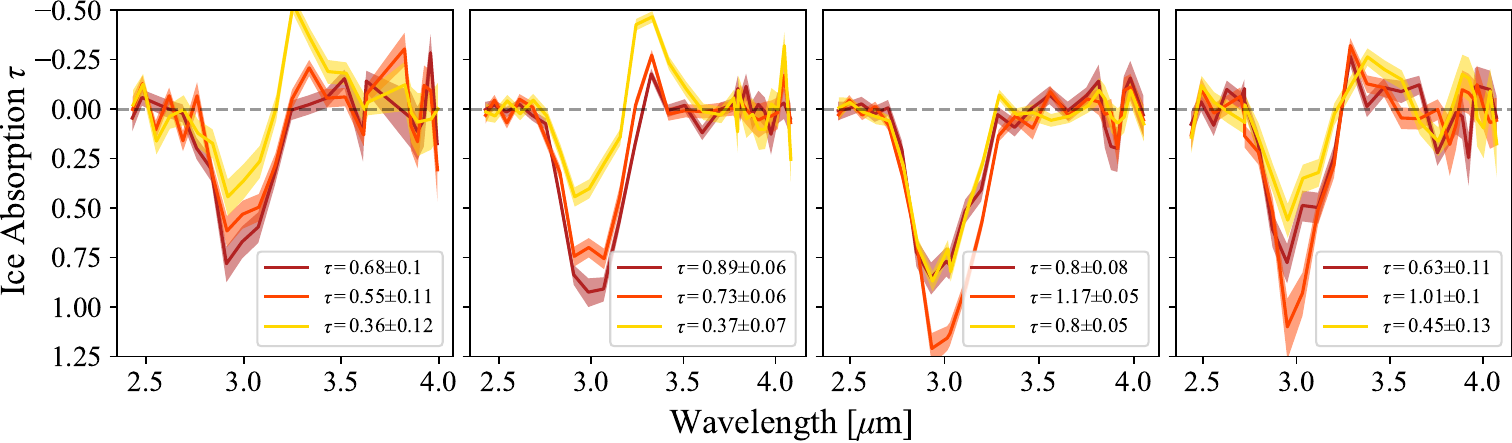}

 \caption{Continuum flux maps and spectra for four prestellar cores. The top row shows the continuum flux at 3.4 $\mu$m from WISE, and contours of $\log_{10}\left(N_{\rm H_2}\right)$ estimated using Herschel. In each core, we pick three representative points marked with crosses, starting at the centre (dark red) and going outwards (orange to yellow), chosen as to avoid intersecting background star sightlines. These points probe decreasing $N_{\rm H_2}$ values as shown by the contours. The middle row gives the spectrum (and $\pm 1\sigma$ uncertainty band) for each of these three points in the same colour as their respective markers, along with the fitted continuum in a dotted line, as described in \S~\ref{sec:observations}. The bottom row shows the optical depth computed with respect to this continuum, where the ice absorption feature strengths can be compared. The full ice absorption maps and spatial trends are shown in Fig~\ref{fig:prestellar-full-2}. Apart from the 3 $\mu$m ice absorption band, we also see the 3.3 $\mu$m PAH emission in L260, L1512 and L1544.
    } 
    \label{fig:prestellar-full-1}
\end{figure*}

This process broadly follows the procedure of \cite{Hora2026}, except the choice for the target region size and the grid resolution/pixel size. We also use more than one point to compute \tauice\ to reduce the effect of noise or unremoved spikes, although this does not affect most pixels.  Figure \ref{fig:prestellar-full-1} shows example flux spectra with quadratic continuum fits, along with the resulting optical depth spectra, for three individual coordinates chosen to avoid stellar sightlines and probe the diffuse scattered flux. The uncertainty bands around the spectra (estimated directly from the flux uncertainties given by SPHEREx) are small compared to the depth of the ice feature. Typical errors in \tauice\ (estimated from Monte Carlo noise realisations of the spectra) are $\leq 15\%$.

Our quadratic fit is usually reliable, making use of $10-12$ points in nearly all pixels in all the sources. Pixels deeper inside each core have more usable continuum channels since the higher extinction renders background stars invisible, so there are no spikes originating from pixels intersecting stellar sightlines. The $S/N \leq 3$ removal primarily affects the outermost regions in each image square, well outside the typical physical extent of the core. 

\begin{figure*}
 \centering
 \includegraphics[width=0.8\linewidth]{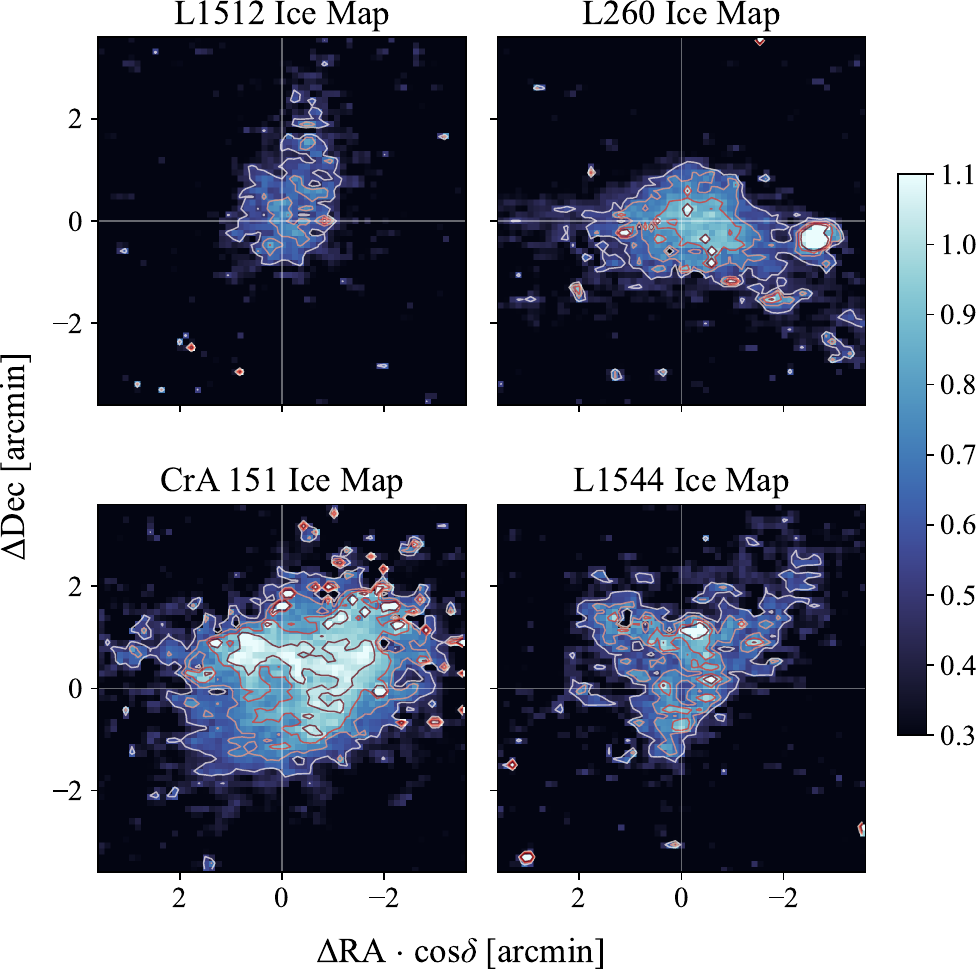}\vspace{15pt} 
 \caption{Ice absorption maps and radial profiles for four prestellar cores. We show the ice absorption depth, $\tau_{\mathrm{H_2O}}$ in a spatial grid, with contours marking the increasing levels of \tauice. 
    } 
    \label{fig:prestellar-full-2}
\end{figure*}

To ensure the robustness of our quadratic continuum fit given the unequal number of points blueward and redward of the H$_2$O ice feature, we also tested the following approaches:
\begin{enumerate}
    \item Change from a quadratic to a linear fit with the same set of points
    \item Compute the median flux in $2.4-2.7$ $\mu$m and in $3.6-4.1$ $\mu$m and fit a line to these two points at the corresponding wavelength midpoints. 
    \item Change the red-side window of the quadratic fit to $3.8-4.1$ $\mu$m, thus avoiding any residual contamination by mixed ice bands of water with ammonia or other species. 
\end{enumerate}
The resulting optical depths change at a $\leq 10\%$ level, and none of the spatial trends discussed in latter sections is significantly affected by the choice of continuum. We thus use the original method of fitting a quadratic to fluxes outside the ice band. 
Doing this for each grid point allows us to construct the spatial \tauice\ maps. The final maps are spatially smoothed with a Gaussian kernel of $\sigma=0.5$ pixel, and are presented in the top panels of Fig~\ref{fig:prestellar-full-2}. Each reference point (RA,Dec) = (0,0) is set to the core coordinates in Table~\ref{tab:core-sample}.

\cite{Hora2026} mention several caveats of using a similar process to create ice optical depth maps. An important distinction (apart from the physical scales) that makes our analysis less prone to those issues is the source of background light. In the \citet{Hora2026} analysis, the diffuse Galactic background is the light source and ice is traced by absorption by the clouds, whereas in our analysis the light source is from scattered light in the cloud itself (coreshine) and it is absorbed by the ice in the intervening portions of the cloud. The diffuse background used in their analysis is considerably fainter than the typical coreshine flux used by us, so their resultant spectra have lower S/N (and more so in the ice absorption trough). This causes saturation of the ice map at \tauice$\sim 1$ due to fluxes below the detector limit in their analysis, whereas we do not find fluxes below the detector limit when removing points with S/N$\leq3$, largely because our cores are significantly illuminated compared to the large-scale backgrounds in these regions. Finally, two more caveats discussed include emission from the surface of the cloud and foreground emission. The cores in our sample are all closer than 150 pc, so that any Galactic foreground flux is likely to be subdominant compared to coreshine (quantitatively discussed in \S~\ref{subsec:foreground}). Further, while scattering (including from the cloud surface) is an important effect here, we explicitly model the cloud in the following section to trace the origin of the continuum light and the ice band absorption. Thermal emission in the NIR can be neglected for these cores as they have typical temperatures of $\lesssim 10-20$ K.

\subsection{Observed Continuum and Ice Maps}
Fig~\ref{fig:prestellar-full-1} shows the continuum maps (from the Wide-field Infrared Survey Explorer, WISE \citealt{Wright2010}, plate scale 2.75 arcsec/pixel, hosted at IPAC \citealt{WISE-IRSA}), raw spectra (from SPHEREx) and fitted continua, and the optical depths for three representative points moving away from the centre of the core. In the top row we also show contours of $\log_{10}\left(N_{\rm H_2}\right)$ in cm$^{-2}$ estimated using Herschel, from the Gould Belt Survey and other programs \citep{Mottram2017, Bresnahan2018, Jensen2024, Kirk2024} hosted on IPAC \citep{Herschel-IRSA}. Fig~\ref{fig:prestellar-full-2} shows the ice absorption depth across the full field of view (pixel size of 7.2 arcsec), along with the radial profile of \tauice.  

In all four cores, we detect a prominent diffuse scattered NIR continuum. We see a prominent polycyclic aromatic hydrocarbon (PAH) emission band around 3.3 $\mu$m \citep{PAH1991} for three of the four cores (except CrA 151), with strength increasing in the outer points as expected. All-sky PAH maps from SPHEREx have also been analysed by \cite{Murgia2026}, and their maps along with our zoom-ins to specific cores constitute the first detection of PAHs in starless cores. 

We also see $\tau_{\rm ice}\approx0.5-1$ in the central region of the cores, which eventually falls to $\approx 0$ well outside the centers. This is primarily due to the continuum flux itself significantly decreasing outside the core when coreshine is no longer efficient, so the ice bands in the spectra no longer satisfy the S/N$\geq3$ criteria. However, visually inspecting a few spectra in these outer regions also shows no significant ice absorption trough, as would be expected for the ambient lower density interstellar medium ($n_{\rm H_2} \lesssim 10^3-10^4$ cm$^{-3}$) where ices cannot easily form. Further, comparing to the $N_{\rm H_2}$ contours of Fig~\ref{fig:prestellar-full-1}, we see ice absorption features predominantly in regions of $N_{\rm H_2} \geq (5-10)\times 10^{21}$ cm$^{-2}$, or $A_V\geq 5-10$. Note that this is different from the $A_V \geq 1.6-3$ threshold for ice formation \citep{Boogert2015} since one is the total line-of-sight $A_V$, and the other is $A_V$ encountered by light incident on the core alone. We also use this value $A_V=5$ to spatially separate a `core' from the larger-scale cloud/filament structure around it. While this is the total line-of-sight column density and cannot be trivially translated to a volume density, this would correspond to a $n_{\rm H_2} \gtrsim (1-3)\times10^4$ cm$^{-3}$ for a typical 0.1 pc core length scale. 

For the first two cores (L1512 and L260), the continuum map shows a flux steadily increasing toward the core centre, while the last two (L1544 and CrA 151) show a central depression in their continuum flux. This effect has been well-studied and modelled in earlier coreshine literature \citep[see for example][]{Steinacker2015}. In the ice maps, the first two cores similarly show an absorption depth increasing toward the centre, apparent both in the full map and in the radial profiles. However, in the latter two cores, the ice absorption is reduced at the centre and instead peaks at intermediate radii. We also check if this can be explained by contributions from foreground flux in Appendix~\ref{subsec:foreground}. We find that this does not change the spatial trend, which therefore is most likely physical. 

An important distinction with the more traditionally used background star measurements is that the continuum flux does \textit{not} originate from behind the core. The core's own scattered light serves as the continuum against which the ice absorption depth is measured. This scattered light can originate from multiple points along the line of sight. This implies that unlike a background source sightline, a simple multiplicative factor cannot be used to derive the total ice column density from the ice depth measured against coreshine. In the next section we develop simple analytical and simulated models to hence probe the origin of the continuum flux and ice absorption within the core.

\section{Modelling-based Insights}
\label{sec:model}

Since the ice absorption maps consist of scattered light originating from the incident interstellar radiation field (ISRF) on the core, we attempt to understand how well the ice absorption optical depth traces the actual ice density in the core, as well as reproduce the observed spatial variations of $\tau$ qualitatively.

To do so, we use both a single-scattering analytical model with a background screen of light, as well as a full 3D Monte Carlo simulation in a 3D isotropic ISRF performed with RADMC-3D \citep{RADMC-2012}. 

\subsection{Physical Parameters}
In both models, we use a common prescription for parameterizing the core structure and properties. We assume a spherical cloud where the density $\rho$ is constant within an inner core (radius $r < r_0$), surrounded by an envelope with the density decreasing as a square-law with radial distance from the centre:

\begin{equation}
\rho(r) = \begin{cases}
\rho_0 & \text{for } r<r_0 \\
\rho_0\left(\frac{r_0}{r}\right)^2 & \text{for } r\geq r_0
\end{cases}
\end{equation}

We use typical physical values for prestellar cores to construct the continuum and ice absorption profiles (as a function of radius) and the simulated images. We choose a power-law index of 2 for the envelope to emulate a Bonnor-Ebert profile, described by a scale radius $r_0$, where the density is constant inside $r_0$ and falls off as $r^{-2}$ outside it \citep{Bonnor1956, Ebert1955}. 

The core radius of interest is set to $0.2$ pc, equal to the typical scale of clouds seen in the NIR images, and slightly larger than core radii found in literature (eg. \citealt{Francesco2007}). The constant density central radius is set to $r_0=0.002$ pc (equal to typical values for these cores; see for example \citealt{Pattle2016, Caselli2019}). We choose different values for the central dust mass density ranging between $\rho_0=4\times10^{-19}$ g cm$^{-3}$ and $\rho_0=4\times10^{-21}$ g cm$^{-3}$. Assuming a gas-to-dust mass ratio of $\sim 100$ \citep{Tricco2017}, this roughly translates to a molecular density range of $n_{\rm H_2} = 10^7$ to $10^5$ cm$^{-3}$ respectively. 

We generate dust grain absorption/scattering opacities and scattering matrices using \texttt{OpTool} \citep{OpTool2021} assuming a DHS dust model \citep{DHSdust2005} with $50\%$ porosity, $\eta_{\rm ice} = 40\%$ ice mantle mass fraction (other values of $\eta_{\rm ice}$ are explored in \S~\ref{subsec:ice-massfrac-variations}) and a standard MRN \citep{MRN1977} power-law size distribution with index 3.5, and grain sizes ranging from a minimum of 0.005 $\mu$m to a maximum of 5 $\mu$m. The upper limit ensures that the dust grains are of a size comparable to the scattered NIR wavelengths used in this work and have a high enough albedo necessary to produce coreshine \citep{Andersen2013, Steinacker2015}. For this fiducial model we do not vary the ice layer thickness and do not include other ice species like CO and CO$_2$. For the Monte Carlo simulation, we use the full absorption/scattering matrix from \texttt{OpTool}.

\subsection{Analytical Model (Single Scattering)}
\label{subsec:analytic-model}

\begin{figure*}
\begin{center}
    \includegraphics[width=0.6\linewidth]{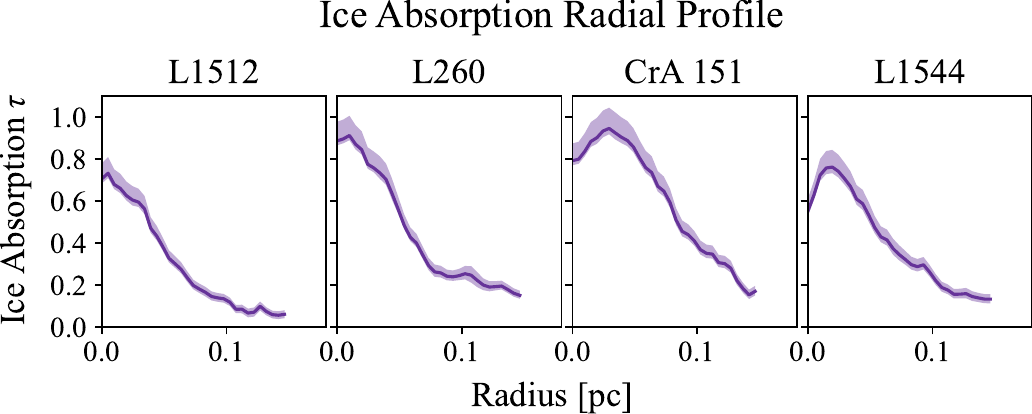}
\end{center}
 \includegraphics[width=6cm]{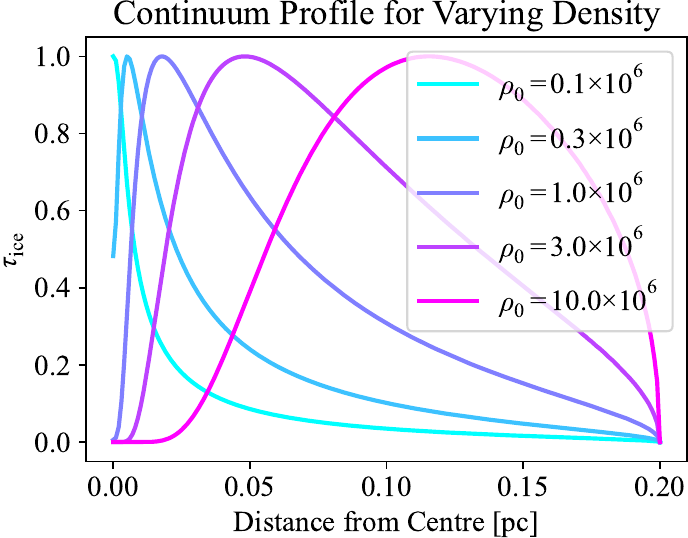}
 \includegraphics[width=6cm]{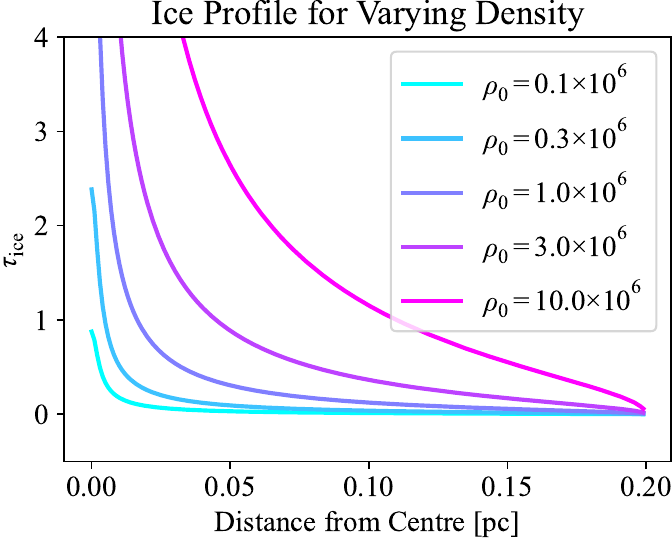}
 \includegraphics[width=6cm]{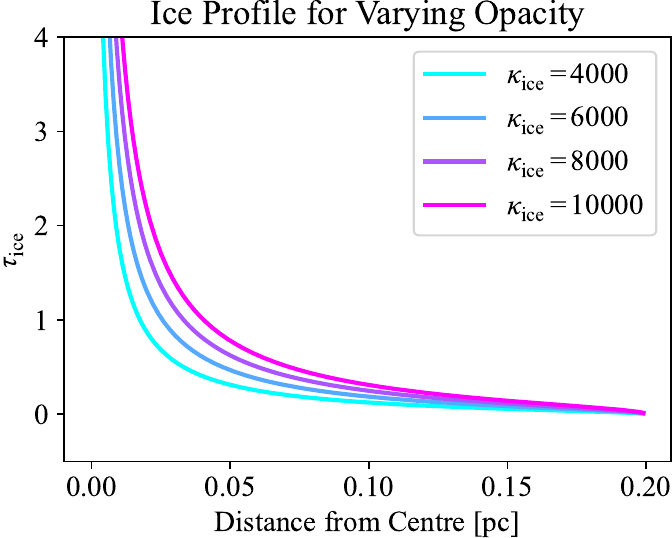}
 \caption{\textit{Top row:} Averaged radial profile of ice absorption depth \tauice\ (and $\pm 1\sigma$ uncertainty band -- upper error including an additional 10\% for the unknown \tauice\ correction factor) observed in the four cores. The lower density cores on the left (L1512 and L260) have a monotonically decreasing ice depth centre to outwards. The higher density cores on the right (CrA 151 and L1544) show a drop in ice absorption depth at the centre. \textit{Bottom row:} Model derived light profile and ice absorption profile as a function of radius for the prestellar core as described in \S\ref{subsec:analytic-model} -- continuum profile for varying densities (left), ice absorption profile for varying densities (middle) and ice absorption profile for varying ice opacities (right).
    } 
    \label{fig:model-pre}
\end{figure*}

We first present the results of a simple analytical model for a spherical cloud illuminated by a uniform background screen, which allows us to efficiently explore how the core physical properties (namely, central density and ice opacity) influence the continuum and ice band spatial trends.  The model setup is presented in Appendix~\ref{appendix:analytical_model}, with results in Fig~\ref{fig:model-pre}.

The left panel of Fig~\ref{fig:model-pre} shows the light profile for the range of chosen central molecular densities. We see that the higher density core reproduces a clear central drop in flux, consistent with CrA 151 and L1544 (the two densest cores) showing a central dim region in NIR continuum in Fig~\ref{fig:prestellar-full-1}. The lower density models do not show any drop in the central flux, which is consistent with the actual low density cores (L260, L1512) \citep{Steinacker2015}. 

The middle and right panels show the variation of ice absorption as a function of radius, for a range of densities and opacities. We see that all of these predict ice absorption to monotonically increase toward the centre of the core. While this is consistent with the observed pattern in the lower density cores (L260, L1512), it is inconsistent with the high density cores where the $\tau_{\rm ice}$ peaks at intermediate radii and then decreases at the centre.

We note here that the setup above is highly simplified, and does not account for any geometrical effects such as the non-spherical nature of most clouds. Further, the scattering is forward-only i.e. the photons do not get deflected in a different direction and continue along the original direction. This approximation is suitable for small-angle scatterings. This is thought to be dominant in the ISM by the Henyey-Greenstein (HG) phase function with $g=0.6-0.8$ \citep{Mathis1990, Gordon2004}. However, more realistic models could incorporate both the anisotropy of the cloud structure and the anisotropy of the ISRF (which can be tuned to be specific to the star-forming region being studied in each case) in analytical models. 

To overcome some of these caveats, we next perform a full 3D Monte Carlo radiative transfer simulation of this similar setup, to check for consistency with the analytical approach, as well as to see if it can explain the anomalous spatial trend of ice absorption in the denser cores. The \texttt{OpTool}-based scattering matrix also self-consistently computes the HG phase function factor $g$, which are hence incorporated in the RADMC-3D models, thus accounting for anisotropic scattering.

\subsection{RADMC-3D Modelling}
\label{subsec:radmc-main-model}
We use the radiative transfer tool RADMC-3D to generate Monte Carlo simulations of photon packets travelling through the cloud in both the NIR continuum band as well as the ice absorption band.

We set up the same spheres as in the previous subsection with two central molecular densities of $10^7$ cm$^{-3}$ and $10^5$ cm$^{-3}$ (the extremes of the range adopted earlier), a central radius of $r_0=0.002$ pc, a power-law profile with index 2 and a truncation radius of 0.2 pc. The cloud is set up on a Cartesian grid of resolution $40\times40\times40$, which we found sufficient to study the trends of light profiles and ice absorption while keeping the computational time reasonable. The dust scattering matrix for the full wavelength range is generated from \texttt{OpTool}. We check for intermediate values of density in \S~\ref{subsec:ice-massfrac-variations}.

We emulate an isotropic ISRF by creating a Fibonacci lattice sphere of $\approx 500$ stars at a radius of 0.4 pc. Their flux density as a function of frequency follows the ISRF prescription in \cite{Mathis1983}. In practice, coreshine modelling attempts use a full Galactic ISRF with directionality \citep{Andersen2013, Steinacker2015}, but that would necessitate separate modelling for each core based on sky positions and is beyond the scope of this paper, as we only seek to reproduce the qualitative behaviour in these cores and not the exact numbers. Further, using a grid of stars instead of a true ISRF allows us to check the effect of removing different parts of the sphere (for example, using a back-hemisphere-only ISRF instead of the full sphere) and examine those effects on the observed light and ice profiles. 

We create images of the cloud for 20 different wavelengths from 2.2 $\mu$m to 4.1 $\mu$m in steps of 0.1 $\mu$m, comparable to the SPHEREx data, using $10^6$ photons at each wavelength. Then we apply the same steps of continuum estimation, normalization and compute the pixel-wise absorption coefficient $\tau$ at the scale of the resolution element. This creates a map similar in format to Fig~\ref{fig:prestellar-full-2} that can be directly compared to see the qualitative spatial trends. We create three maps, one each with viewing axis along $x-, y-$ and $z-$axis of the cloud to minimise the effect of directionality in viewing. We also check for the effect of a non-isotropic ISRF by limiting the stellar screen to only the back hemisphere (back relative to the line-of-sight) and did not find any significant differences, as would be expected for forward-dominated scattering.

\begin{figure*}
\centering
 \includegraphics[width=6cm]{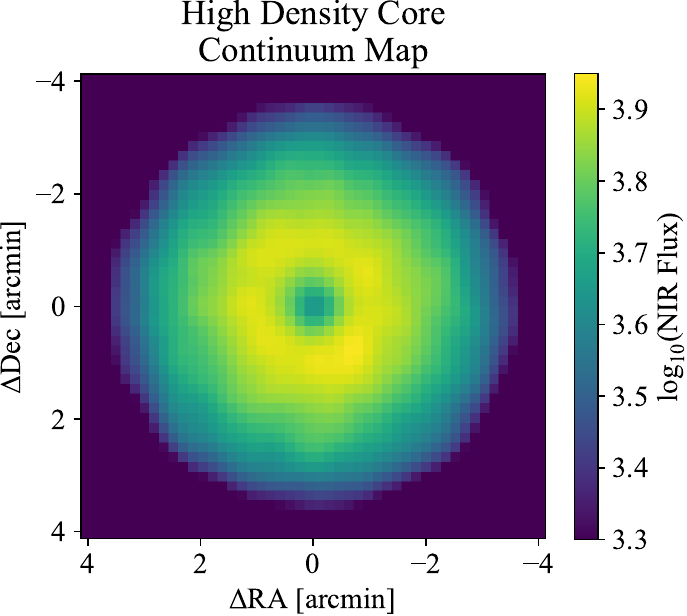} \hspace{1.2cm}
 \includegraphics[width=6cm]{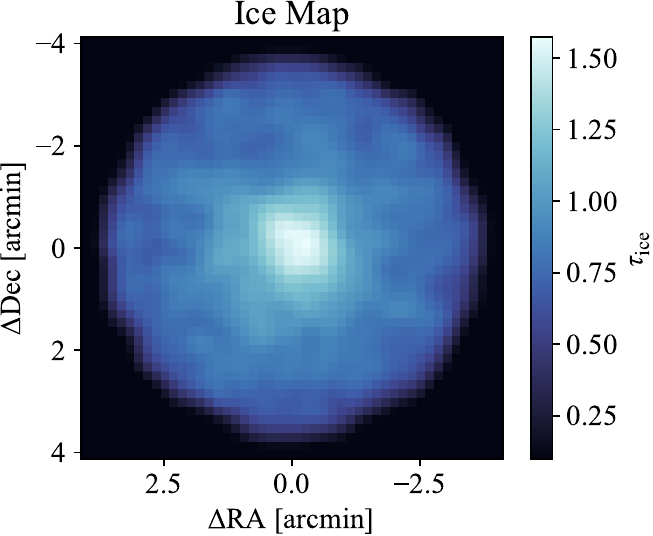}
 \\ 
 \includegraphics[width=6cm]{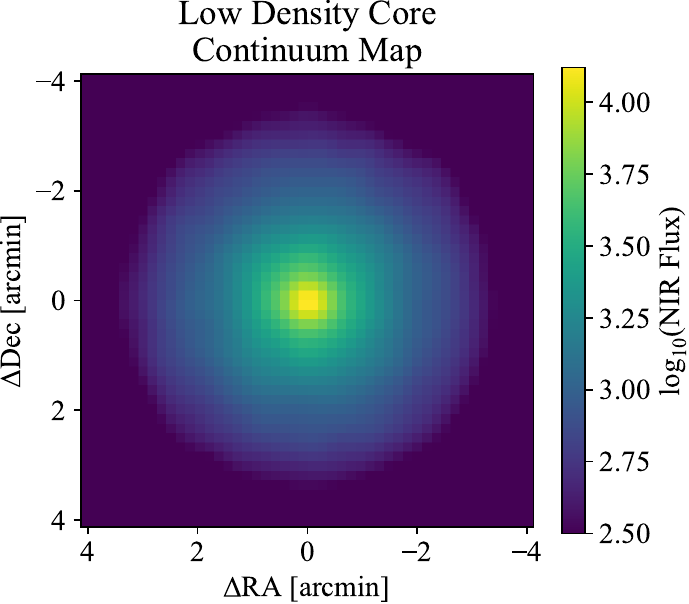} \hspace{1.2cm}
 \includegraphics[width=6cm]{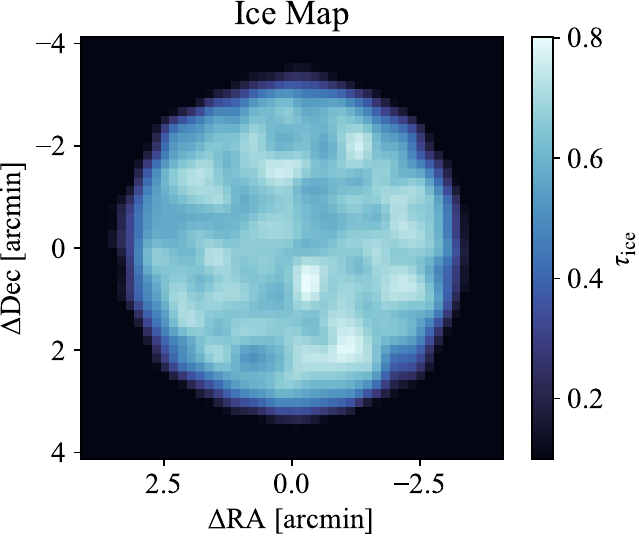}
 \\ 
 \caption{{\it Top row}: RADMC-3D derived NIR light profile (left), observed ice absorption profile (right) for the dense (central $n_{\rm H_2}\approx 10^7$) prestellar core. The central flux dip is reproduced in NIR continuum. We observe the central ice absorption to be very high. All profiles are roughly consistent with the analytical model in Fig~\ref{fig:model-pre}. {\it Bottom row}: RADMC-3D derived light profile and ice absorption profile for a lower density core (central $n_{\rm H_2}\approx 10^5$). 
    } 
    \label{fig:radmc-pre}
\end{figure*}

We present the NIR continuum light profile and the ice absorption map for each of the two central densities in Fig~\ref{fig:radmc-pre}. We adopt an angular scale of the 0.2 pc cloud at a distance of 140 pc for direct comparison to the maps in Fig~\ref{fig:prestellar-full-1}~and~\ref{fig:prestellar-full-2}. Typical surface brightness are $10^3-10^4$ megajansky per steradian, and we set a flux lower limit of 200 on all pixels at all wavelengths to avoid noise artifacts (the scaling difference is due to the ISRF luminosity normalisation; \tauice\ calculations use ratios and hence remain same). 

The NIR images and ice absorption maps are qualitatively similar to the analytical (radial) profiles.  The low-density model matches the observed maps in \S~\ref{sec:observations} for L260 and L1512, with a centrally peaking continuum flux and relatively flat ice absorption profile. The differences in the exact values of $\tau_{\rm ice}$ are about $20-40\%$, which can be attributed to using a somewhat oversimplified cloud model for this scenario.  For the higher-density model, we again see a central dip in the continuum flux but still do not see a central decrease in ice absorption as is seen in the SPHEREx maps of CrA 151 and L1544.

\subsubsection{Simulated Model Insights}
\label{subsec:ice-massfrac-variations}
To check if the observed \tauice\ can reasonably distinguish between different total ice column densities, we construct the radial profile of \tauice\ by averaging along annuli at varying radii in the ice map, for the following two scenarios:
\begin{enumerate}
    \item Holding the ice mass fraction $\eta_{\rm ice}=0.4$ constant, and varying the central density $n_{\rm H_2}$ to five values between $10^5$ and $10^7$ cm$^{-3}$
    \item Holding the central density constant $n_{\rm H_2} =10^6$ cm$^{-3}$ and varying the ice mass fraction $\eta_{\rm ice}$ between 0.1 and 0.9.
\end{enumerate}

\begin{figure*}
\centering
 \includegraphics[width=6.1cm]{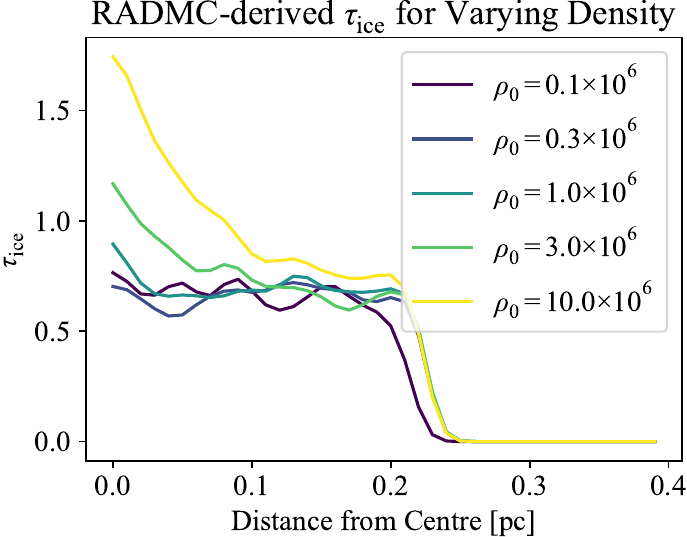} \hspace{1.2cm}
 \includegraphics[width=6.5cm]{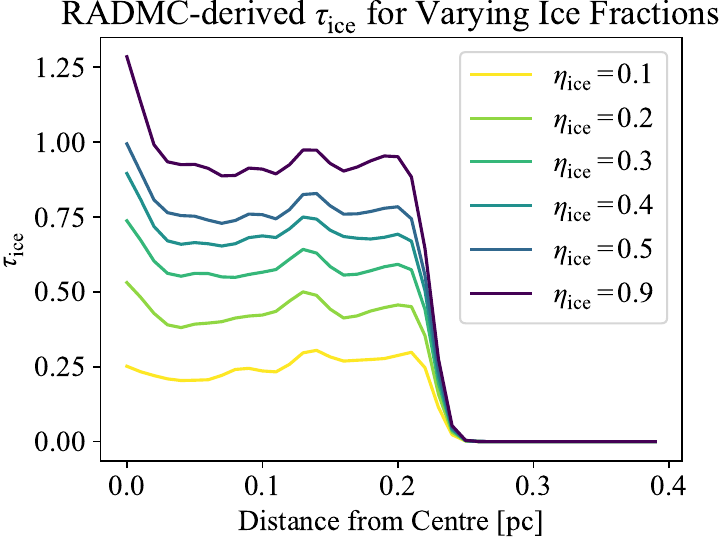}
 \\ 
 \caption{{\it Left panel}: RADMC-3D derived \tauice\ as a function of radius for varying central densities. {\it Right panel}: RADMC-3D derived \tauice\ as a function of radius for varying ice mass fractions $\eta_{\rm ice}$. We see a clear difference in the observed \tauice\ profile as $\eta_{\rm ice}$ is varied.
    } 
    \label{fig:radmc-radial}
\end{figure*}

The plots for both scenarios are shown in Fig~\ref{fig:radmc-radial}. The variation in \tauice\ is less evident with changing central densities, especially in the outer regions. The innermost regions show some sensitivity to the density, but these would be reliably distinguishable only for the highest density values. This makes it less useful as a probe of the absolute gas/dust density at a fixed ice fraction.

However, the absolute density structure of the cloud is often inferred through dust continuum emission \citep[for example][]{Launhardt2013, Konyves2015, Kirk2024} or gas-phase line emission such as from CO \citep{Sanhueza2019, Koley2025}. The unknown then is the ice density or column, which is a proxy for the ice mass fraction at a fixed dust density. The right panel in Fig~\ref{fig:radmc-radial} shows that \tauice\ can reliably probe the ice mass fractions for a reasonable range of $\eta_{\rm ice}$ between 0.1 and 0.9. The set of models shows that if the geometry and central density is known (specifically $\rho_0$ and $r_0$ of the best-fitting Bonnor-Ebert sphere), we can constrain the ice mass fraction (and hence the ice column density) from coreshine-derived spectra. 

Such mapping thus does not necessarily require background sources. The two lower-density cores (L260 and L1512) show \tauice\ values of $0.8-1$, fully consistent with the model-derived values for their known central $n_{\rm H_2}$ and expected $\eta_{\rm ice}$ ranges of $\sim 0.2-0.9$. 

Note that due to the high total line-of-sight opacity (\tauice$>20$) through the central regions in the densest clouds (Table~\ref{tab:core-sample}), the scattering-based \tauice\ value primarily probes the front side of the cloud in the central regions of the observed map. It does not probe all the way to the cloud centre, hence necessitating the above modelling step. This is clear from the RADMC-3D central \tauice\ values of 1--2, far lower than the total line-of-sight value of 30. So we do not interpret the observed \tauice\ as a direct measurement of the central ice abundance.

However, for a fixed density profile under the standard assumption of a constant $\eta_{\rm ice}$, the central ice density is proportional to the ice density on the `front side'. Hence what Fig~\ref{fig:radmc-radial} shows is that \tauice\ can measure the front side ice density. It can hence indirectly measure the central ice density if $\eta_{\rm ice}$ is fixed, or measure $\eta_{\rm ice}$ if the central density/front-side density is fixed. The \tauice\ profiles and values directly constrain the front-side ice density; however if the true central ice density (in the \tauice$\gg 1$ region) departs from the constant ice fraction assumption, our scattered light maps will not be able to determine that.

While exact values of the ice column density cannot be calculated directly as for background star sightlines, it should be possible to infer using modelling if the density structure is well-constrained from dust continuum or gas-phase measurements. Further, since roughly half of known cores show coreshine \citep{Pagani2010}, this technique combined with SPHEREx all-sky mapping can be used to study spatial ice profiles in a large number of cores.

However, these models cannot reproduce the observed dip in ice absorption toward the higher density cores, instead showing an increase or plateauing in the \tauice\ value close to the centre for the full range of central densities and ice mass fractions adopted. 
\\ \\
\noindent As mentioned earlier, these models are intentionally simplified: they do not account for anisotropies in either the source of illumination or the cloud structure, do not include the effects of the physical and chemical diversity of dust grains, and do not capture dependencies on the surrounding galactic environment and the parent molecular cloud. Also, we do not vary the physical parameters over a full phase space of possible values for the RADMC-3D simulations and stick to the fiducial values used in the primary analytical model. While the results from the 3D Monte Carlo simulation and the analytical model broadly agree, exactly matching the models to the observations of $\tau_{\rm ice}$ would require a detailed simulation suite including the aforementioned effects and over a larger parameter space. This is beyond the scope of this work and we only seek to test whether the qualitative spatial trends and approximate values for $\tau_{\rm ice}$ can be reproduced with the simple models constructed above.  

\section{Discussion}
\label{sec:discussion}
While L1512 and L260 show a NIR flux that uniformly decreases from central to outer regions, L1544 and CrA 151 show a central flux depression. Detailed models of galactic coreshine \citep{Pagani2010, Steinacker2015} self-consistently produce this central depression from scattering in a dense core with high central densities. Our analytical model and Monte Carlo simulation both produce the same behaviour for models with high central densities ($n_{\rm H_2}=10^7$ cm$^{-3}$). This is consistent with observationally measured central densities in both CrA 151  and in L1544 ($\rho_0=10^7$ cm$^{-3}$, \citealt{Redaelli2025, Keto2014}). On the other hand, L1512 and L260 both have $\rho_0\approx10^4-10^5$ cm$^{-3}$ \citep{Lin2020, Steinacker2015, Jensen2024}, and our models predict a uniformly increasing flux toward the centre at those densities. 

While the NIR continuum fluxes have been extensively studied both theoretically and observationally, SPHEREx now enables the creation of ice maps on the same physical scales. In these, we see a similar diversity in the ice profiles -- the two lower-density cores show increasing ice absorption depth toward the centre, while the ice absorption toward the denser cores appears to flatten or decrease toward the centre. Since spatially resolved absorption maps were not widely accessible prior to SPHEREx, this trend within a single prestellar core has not been observed prior to this.

Our models do not reproduce the shallower central ice absorption in the densest cores, when assuming the ice density and absorption scales linearly with the gas density, in a simple isotropic ISRF with unchanging dust parameters. Further, while our analytical models use a backlit ISRF, the RADMC-3D models use a spherically isotropic ISRF. For the optically thin outskirts, such diffuse background Galactic light could trace the full line-of-sight ice column (similar to Cygnus-X in \citealt{Hora2026}) and increase observed \tauice. Meanwhile the optically thick centre only has scattering from its frontmost regions giving a lower \tauice. We attempt to show the \tauice\ map in this Galactic-background-only case in \S~\ref{subsec:isrf-alt-geometry}. However, this scenario needs more careful modelling, including realistic relative brightness contributions of the Galactic plane and the surrounding star-forming cloud in each case and cannot be ruled out with our present models. In either case, this effect does not fully account for the inconsistency in the absolute value of central \tauice\ in the optically thick regime seen in models (1--1.8) and the observed maps (0.6--0.8).

We now aim to explore the effects of alternative geometries, possible grain growth in the central regions, or density scaling relations of ice with gas in our models, and check if any can explain this spatial pattern. For all scenarios, we use RADMC-3D simulations rather than the analytical model since this provides more realistic scattering physics. A central density $n_{\rm H_2}=10^7$ cm$^{-3}$ was adopted for all the tests, since the unexplained spatial trend occurs in the higher density cores.

\begin{figure*}
\centering
 \includegraphics[width=6cm]{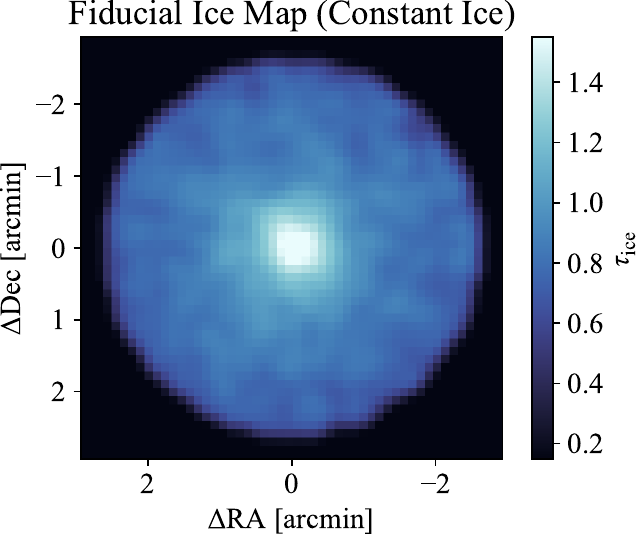} \hspace{14pt}
 \includegraphics[width=6cm]{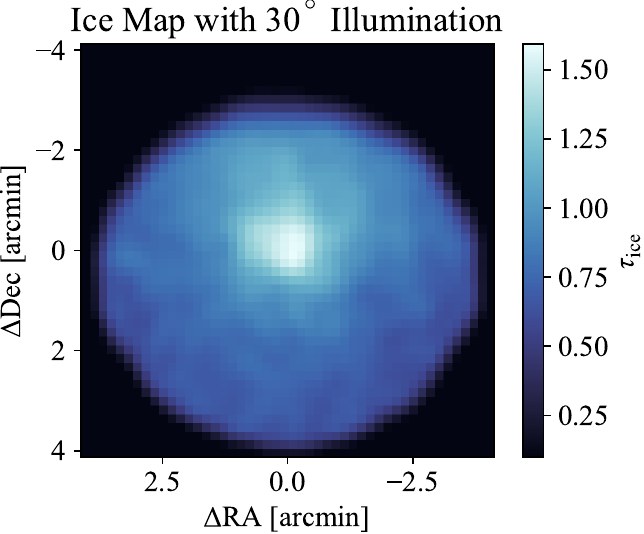} \vspace{10pt}\\
 \includegraphics[width=6cm]{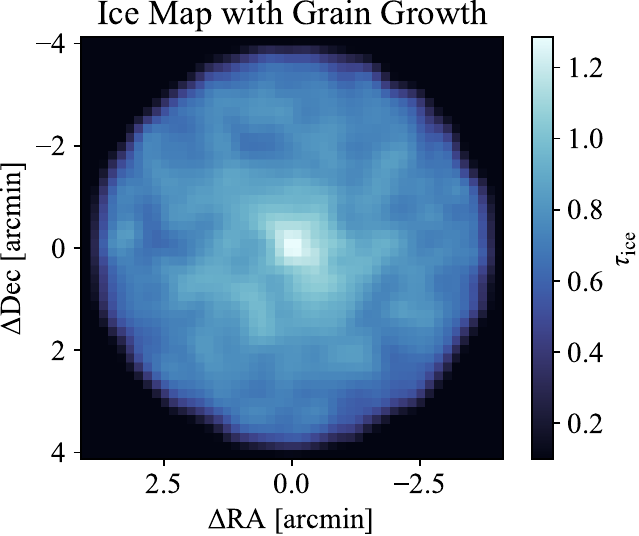} \hspace{14pt} \includegraphics[width=6cm]{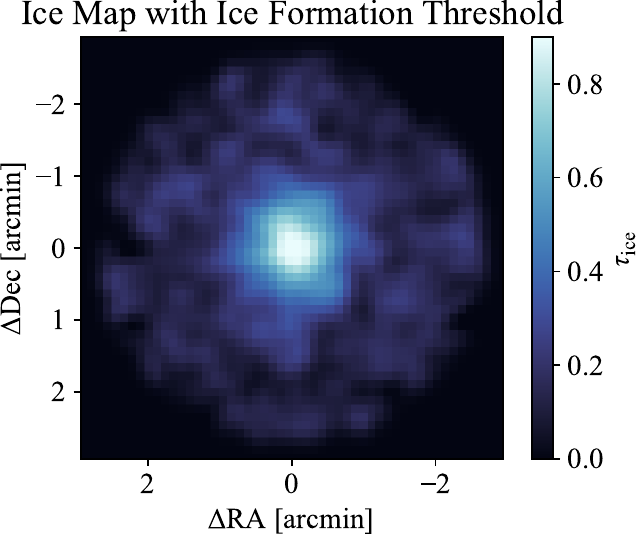}
 \vspace{10pt}\\
 \includegraphics[width=6cm]{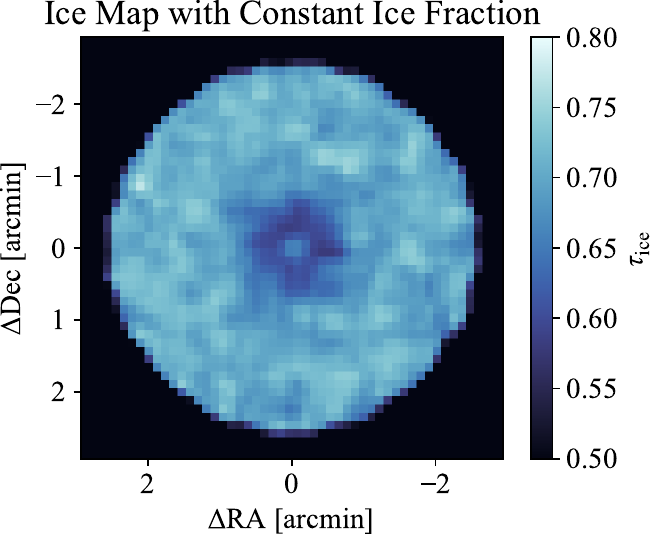} \hspace{14pt} \includegraphics[width=6cm]{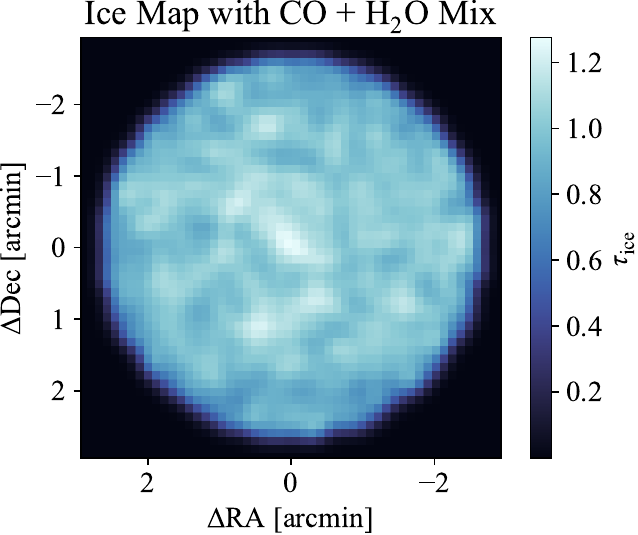}
 \vspace{8pt}
 
 \caption{RADMC-3D derived ice absorption map for \textit{Top left}: the fiducial cloud for a constant $\eta_{\rm ice}=40\%$, same as Fig~\ref{fig:radmc-pre}, for comparison {\it Top right}: a cloud illuminated at $30^\circ$ to the line-of-sight, the source of illumination being in the top part of the image. \textit{Middle left}: a cloud where central regions have grain sizes reaching up to 100 $\mu$m instead of 5 $\mu$m. 
 \textit{Middle right}: a cloud where ice formation primarily happens at densities $n_{\rm H_2}\geq10^4$ cm$^{-3}$. \textit{Bottom left}: a cloud where ice density becomes constant/saturates above gas densities of $n_{\rm H_2}\geq10^4$ cm$^{-3}$, and does not scale with the increasing gas/dust density.
 \textit{Bottom right}: a cloud where ice changes from being pure H$_2$O, to a mix of a 3:1 mix of CO and H$_2$O ice, at densities $n_{\rm H_2}\geq10^4$ cm$^{-3}$. 
    } 
    \label{fig:alt-models}
\end{figure*}

\subsection{Alternative Geometry}
\label{subsec:isrf-alt-geometry}
Most of these clouds are located at Galactic latitudes of $10^\circ-20^\circ$, and if the Galactic plane was the primary source of illumination, one would expect similar ($10^\circ-20^\circ$) scattering angles to dominate the NIR flux seen along our LoS. However, these clouds are all located in regions of active star formation, and it could be that the local neighbourhood is the dominant source of incident radiation, and is thus closer to isotropic.

To test the importance of the incident radiation field angle, we run a new model in RADMC-3D where a spherical cap-shaped ISRF is used instead of an isotropic ISRF. The half angle of the cap was set to $30^\circ$, but the exact number does not make a difference in the results. 
This cap is angled at $30^\circ$ to the LoS, implying the dominant scattering seen in observed images is due to light scattered at $\sim30^\circ$. We no longer average over sightlines along three axes since the angular dependence is the factor being tested. 

The resultant ice absorption map is shown in the top right panel of Fig~\ref{fig:alt-models}. The asymmetry is immediately clear, and an effect like this may explain the asymmetric profile in CrA 151 (Fig~\ref{fig:prestellar-full-2}), where the north-west (top right) part of the cloud is brighter than the south-east. However, the central peak in $\tau$ is still present, demonstrating that the radiation geometry cannot explain the observed dip in absorption. We next test whether changing opacities due to dust grain growth can cause the observed dip in ice absorption.

\subsection{Grain growth in central regions}
\label{subsec:large-grains}
Dust grains are expected to coagulate from smaller to larger sizes as the density of gas in core increases \citep{Ormel2009}. Hence, for denser cores with central $n_{\rm H_2} \approx 10^7$ cm$^{-3}$, physical conditions may allow the dust grains to grow beyond the maximum size of 5 $\mu$m assumed so far in all our models.

Dust grains upto $\sim 100-1000\ \mu$m have been used to explain observations in the dense cores, such as in the Orion Molecular Cloud \citep{Miettinen2012, Schnee2014, Nozari2025}, and in envelopes of early stages of protostars \citep{Galametz2019, Caciapuoti2023}. Larger grain sizes can significantly change the continuum absorption and scattering opacities, which can heavily vary with grain sizes. The water absorption band however occurs at 3 $\mu$m band for all grain sizes and changes in a different manner than the continuum opacity. To check if these changes in opacities can drive the observed \tauice\ such that we observe a reduction at the centre, we run the same RADMC-3D model as \S~\ref{subsec:radmc-main-model} but by changing the maximum dust grain size of the MRN distributionfor the central region in \texttt{OpTool} to 100 $\mu$m, i.e. using $dN/da \propto a^{-3.5}$ with $a_{\rm max}=100\ \mu$m.

For simplicity, we create a sharp boundary for the transition in dust composition i.e. for radii $<r_{\rm growth}$, the dust entirely consists of grains with maximum size of 100 $\mu$m, and for radii $>r_{\rm growth}$, the dust has a maximum size of 5 $\mu$m as before. We set $r_{\rm growth}$ to various values of 0.01, 0.03, 0.06 and 0.1 pc. In all cases the central \tauice$\approx1.2$ reduced slightly from the central \tauice$\approx1.5$ in the primary models in \S~\ref{subsec:radmc-main-model}, but none of them could reproduce a scenario where the central \tauice\ is lower than that at surrounding intermediate radii. The ice map for $r_{\rm growth}=0.06$ pc is shown in the middle left panel of Fig~\ref{fig:alt-models}. This shows that even with grain growth to sizes as large as 100 $\mu$m, the observed dip in absorption cannot be fully explained. We next test the effect of different density scalings.

\subsection{Ice formation threshold model}
\label{subsec:alt-models}
Since water ice only forms beyond a certain dust density/extinction threshold (typically $A_V=1.5-3$, \citealt{Boogert2015}), we next test a model where the ice-to-gas ratio increases toward the centre instead of remaining constant as in the preceding section.

\begin{figure}
    \centering
    \includegraphics[width=\columnwidth]{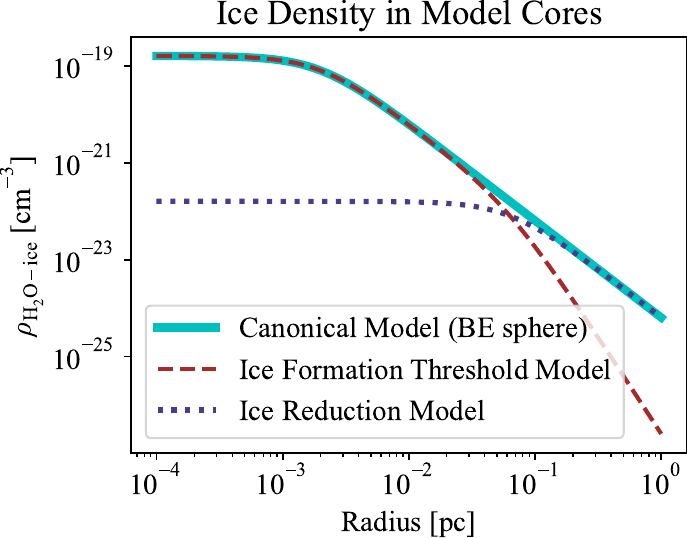} \\
    \includegraphics[width=\columnwidth]{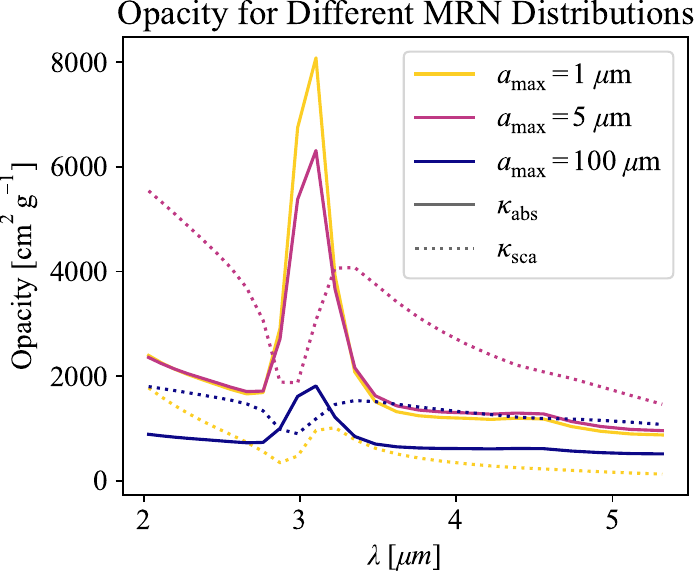}
    \caption{\textit{Top:}Ice density as a function of radius for the canonical model in \S~\ref{subsec:radmc-main-model} in cyan, and for the ice formation threshold and ice reduction models from \S~\ref{subsec:alt-models} in brown and indigo respectively. \textit{Bottom:} Opacity for our wavelength range of interest (2--5 $\mu$m) for typical grain size distributions used in ISM (MRN with $a_{\rm max}=1\ \mu$m), for our canonical model grain size distribution ($a_{\rm max}=5\ \mu$m) and for the central large dust size distribution in \S~\ref{subsec:large-grains} ($a_{\rm max}=100\ \mu$m)}
    \label{fig:ice-radius}
\end{figure}

We introduce a density threshold $\rho_t$ below which the ice density falls off more rapidly than the dust density. Let $f$ be a factor given by 
\begin{align}
    f = \frac{\rho}{\rho+\rho_t}
\end{align}
such that the effective ice absorption coefficient scales as
\begin{equation}
    \kappa_{\mathrm{abs,ice,eff}} = f\cdot\kappa_{\mathrm{abs,ice}} + (1-f)\cdot\kappa_{\rm abs, cont}
\end{equation}
This broadly means that for large densities (closer to the centre, $\rho>\rho_t$), $f \approx 1$ and the ice density scales linearly with the dust density. But for small values ($\rho<\rho_t$), $f \approx \rho/\rho_t$, meaning the ice-to-gas ratio scales as $\rho$, and the ice density falls off faster than the dust density. The resultant density as a function of radius is shown in Fig~\ref{fig:ice-radius}.

We note that the choice of the factor $f$ is arbitrary and used only for the demonstration here, aside from the condition that $f\approx1$ needs to hold at high densities. We choose a density threshold $\rho_t$ corresponding to number densities of $n_{\rm H_2}=10^4$ cm$^{-3}$. For a typical $0.1$ pc length scale and adopting the $N_{\rm H_2}/A_V$ relation from \cite{Bohlin1978}, this corresponds to $A_V\sim3$.

Since RADMC-3D does not natively allow changing the opacities $\kappa$ directly, we create two dust species in \texttt{OpTool}, one with $40\%$ ice covering by mass and one with $1\%$ ice. We start with the base radial density profile $\rho(r)$ of the $n_{\rm H_2}=10^7$ cm$^{-3}$ central density BE cloud. Then we define two density structures as 

\begin{align}
    \rho_\mathrm{40\%\ ice} &= f\rho(r) =\left(\frac{\rho(r)}{\rho_t+\rho(r)}\right) \cdot\rho(r)  \\
    \rho_\mathrm{1\%\ ice} &= (1-f)\cdot\rho(r) 
\end{align}

This is mathematically equivalent to adjusting the absorption coefficient. The resultant ice absorption map from running this setup on RADMC-3D is shown in the middle right panel of Fig~\ref{fig:alt-models}. As is expected, the ice absorption is prominent only toward the centre. This is not in complete agreement with the SPHEREx maps and might point to ice formation even below the standard $A_V>3$ threshold. This could however be due to a larger core radius $r_0$ than what we assume in the models, as the density scales as $(r/r_0)^{-2}$ for a fixed $\rho_0$. In either case, it cannot explain the observed trend of the central \tauice\ being lower than that at its surrounding radii. 

\subsection{Ice Reduction model}
We now reverse the prior ice threshold scenario so that the ice density reaches a saturation threshold in the innermost regions. For the same prescription of $f$, we now use

\begin{align}
    \rho_\mathrm{40\%\ ice} &= (1-f)\cdot\rho(r) =\left(\frac{\rho_t}{\rho_t+\rho(r)}\right) \cdot\rho(r)  \\
    \rho_\mathrm{1\%\ ice} &= f\rho(r) 
\end{align}

Now for large central densities, $\rho(r)>\rho_t$, the ice density $\rho_\mathrm{40\%\ ice}$ becomes a constant $\approx \rho_t$. This means that while dust densities continue to grow, the ice density remains constant. The resultant density profile as a function of radius is shown in Fig~\ref{fig:ice-radius}.

The ice absorption map for this scenario is shown in the bottom left panel of Fig~\ref{fig:alt-models}, revealing a possible central reduction in ice absorption (except at the very central pixel). This shows that one way to reproduce the observed spatial trend is if the ice abundance relative to dust is sufficiently reduced. We explore this possibility in more detail below.

Ice density is generally treated as scaling linearly with column density \citep[see][and references therein for an overview]{Boogert2015} with no evidence for turnover for $A_V$ values up to 90 for background star sightlines \citep{JWSTref3} and dense protostellar envelope sightlines \citep{Pontoppidan2004}. We inspect the $A_V$ derived from $N_{\rm H_2}$ contour maps (using the relation of \citealt{Bohlin1978}) made using Herschel in Fig~\ref{fig:prestellar-full-1}. We find that the ice absorption turnover (i.e. the point where absorption starts becoming shallower again) is at $A_V=25-30$ mag in our maps. Although for these near-central regions \tauice\ increases much beyond 1 so scattered light is not expected to directly probe the central ice (\S~\ref{subsec:ice-massfrac-variations}), we do see an emergent difference in the maps in Fig~\ref{fig:alt-models}. This reduced ice is in tension with previous observations of ice density scalings making it physically unlikely, and may be reconciled with more refined modelling in future work.

This extinction range also corresponds to a flattening of ice absorption seen in background star ice mapping of the Chamaeleon I molecular cloud \citep[Extended Data Fig. 4 of ][]{Smith2025}, although this could be due to instrument sensitivity to high extinction lines-of-sight.

\subsubsection{Oxygen Reallocation in Other Ice Species}
A moderate reduction in water ice density could potentially be caused by the reprocessing of the oxygen into other species within the highest-density inner regions of the cores. Beyond $n_{\rm H_2} \gtrsim 10^4$ cm$^{-3}$, species such as CO rapidly freeze out onto dust grains and dominate the overall ice composition \citep[for example][]{Redman2002, Caselli2022}, which could influence the water ice absorption. While CO and CO$_2$ ice can be probed with SPHEREx as well, they require more careful corrections based on the SPHEREx bandpasses on a pixel-by-pixel basis \citep{Hora2026}, which we defer to future work. 

Water ice absorption was also studied in L1544 along three background star sightlines \citep{Goto2021}. We compare SPHEREx spectra and \tauice\ measurements of the same stars and find good agreement (details and figures in Appendix~\ref{subsec:bgstars}).  The three stars validate our scattered-light \tauice\ measurements for the $A_V$ range up to $\sim 30$ to intermediate radii in the core. However they do not intercept the central densest region, so our higher $A_V$ \tauice\ values (where the discrepancy is) are not directly comparable.

To study the expected effect of oxygen reallocation, we construct a model where the inner regions (at densities $n_{\rm H_2} \gtrsim 10^4$ cm$^{-3}$) are composed of dust grains with $\eta_{\rm H_2O-ice}=10\%$ and $\eta_{\rm CO-ice}=30\%$, instead of $\eta_{\rm H_2O\ ice}=40\%$. The simulated \tauice\ map is shown in the bottom right panel of Fig~\ref{fig:alt-models}.  This is a more extreme scenario than the typical expectation of $\eta_{\rm CO-ice} \lesssim \eta_{\rm H_2O-ice}$, but we still see that \tauice\ plateaus toward the centre but does not reduce. 

Gas-phase molecular abundances have also been well-studied in L1544, including the first detection of gasesous H$_2$O in a prestellar core \cite{Caselli2012} using Herschel. This was subsequently modelled in L1544 by \cite{Keto2014} and \cite{Vasyunin2017}. Gas-phase abundance reduces toward the centre, typically ascribed to freeze-out into H$_2$O ice. The \cite{Vasyunin2017} model predicts a linear scaling of ice and hydrogen density. Thus a drastic drop as proposed in the ice reduction model is inconsistent with predictions. There is still a possibility that water gets processed into more complex molecules in dense regions (proposed by \citealt{Whittet2010} and \citealt{Jenkins2009} to explain an oxygen deficit in the ISM). However, these other species are expected to be at most a few \% as abundant as water ice, and are unlikely to account for the majority of the oxygen budget. 
\\ \\
Hence, despite its qualitative success in explaining the central dip in ice absorption, a drastic reduction in water ice to near-zero levels at the centre still remains difficult to reconcile with physical expectations (as water is still expected to be most abundant among all ice species). Other explanations for the observed spatial pattern remain open. More sophisticated treatments of the geometry of the cloud and ISRF (both from the Galaxy and from the surrounding star formation), the physical and chemical properties of dust and other effects prevalent in high-density cores could self-consistently explain the central decrease in \tauice\ without needing strong modifications in density profiles.

\vspace{5pt}
\subsection{Tracing Changes in Grain Structure and Chemistry with Coreshine}

The models so far have difficulty explaining the observed spatial trend. However, we note that all panels except the top two of Fig~\ref{fig:alt-models} explore scenarios where the high-density conditions in prestellar cores cause the dust grain physics or ice compositions to change from the lower-density outskirts. Comparing the absorption maps in the simulated images, each scenario generates a unique combination of spatial variations and absolute values of \tauice.

We can thus use the uniform spatial ice mapping using coreshine flux as a probe to study the evolution of dust grains and ices in dense cores. Combined with the results of \S~\ref{subsec:ice-massfrac-variations}, this would work as a powerful new tool to probe both ice column densities and dust and ice properties in prestellar cores. Future studies with statistically large samples of cores showing coreshine, as well as more detailed modelling setups, are instrumental in maximising the utility of this physical phenomenon and SPHEREx's all-sky coverage to learn more about dust in protostellar and protoplanetary environments.

\section{Conclusions}
\label{sec:conclusions}
The main outcomes of this work are as follows:
\begin{enumerate}
    \item \textit{We demonstrate the ability of SPHEREx spectrophotometry of near-IR scattered flux in star-forming clouds (coreshine) to trace H$_2$O ice in star-forming cores}. Both our observations and the RADMC-3D modelling show that the ice absorption trough arises in the coreshine spectra around 3 $\mu$m and can reflect spatial variations of ice abundance (albeit the relationship depends on the local scattering geometry and density). In this work, we analyse four cores that are close to Earth and prominent NIR scatterers. We use the high spatial resolution to create maps of ice absorption coefficient $\tau_{\mathrm{H_2O}}$ (with uncertainties $\lesssim 10\%$) across the central $\sim 0.2$ pc at a resolution of $\approx 900-1200$ AU. 

    \item \textit{Our analytical and RADMC-3D models show that the maps probe ice densities in the core for a given geometry.} Over a range of central $n_{\rm H_2}=10^5-10^7$ cm$^{-3}$ and ice mass fraction $\eta_{\rm H_2O}=0.1-0.9$, we show that the RADMC model-derived ice profiles can reliably distinguish between different $\eta_{\rm H_2O}$ (and hence constrain the ice column density), if the overall dust density and geometry of the cloud are well-constrained. Although, for the densest cores, scattering only probes ice till the \tauice$\sim1$ photosphere-like surface. This indirectly probes the central density if the whole core follows the assumption of a uniform $\eta_{\rm H_2O}$. It does not otherwise directly constrain ice column in the \tauice$\gg 1$ regions.
    
    \item \textit{We see evidence of reduced central ice absorption specific to high-density, dynamically evolved prestellar cores, which is inconsistent with simple linear scaling of ice density with gas density.} The two cores with high central densities, L1544 and CrA 151, both show a central dip in continuum flux (explored and modelled in prior literature) along with a shallower ice absorption feature compared to the outer regions, evident in both spatial maps and individual point spectra. We explore if this could be caused by foreground flux addition in the spectra, and find it is most likely physical (barring measurement uncertainties). This cannot be reproduced by a simple Bonnor-Ebert sphere with ice density scaling directly as the dust (and hydrogen gas) density. 
    
    \item \textit{The physical properties of dust grains and compositions of ice species can create observable differences in the \tauice\ maps.} While simple geometric effects, grain growth at high densities and ice formation thresholds cannot reproduce reduced absorption at the centre, these effects can be traced by the spatial patterns and absolute values of \tauice\ in the absorption maps. On the other hand, one potential explanation for reduced central water ice absorption is found by modelling water ice density to reach a constant density/saturate at an intermediate radius and not increase inward of that. This could be partially explained by oxygen locking up in CO ice and more complex (potentially carbonaceous) molecules in the innermost dense regions, reducing H$_2$O ice abundance, but is largely unexpected in light of previous studies of ices towards background stars and dense protostellar envelopes. Other physical explanations cannot be ruled out. For example, both grain growth and inclusion of CO ice in the innermost dense regions, changes the opacities and reduces central \tauice\ compared to the fiducial model (although not below \tauice\ at intermediate radii). Thus the possibility of more complex dust physics and ice species compositions driving this spatial trend remains open.

\end{enumerate}
This preliminary analysis of four cores shows that coreshine can be used as a novel and robust probe to constrain water ice abundances from spectroscopy of diffuse scattered flux.
The exquisite spatial resolution and areal coverage further make it the ideal probe to analyse multiple ice species in future studies, allowing us to study their abundances and dependence on the physics of the local and global environments at physical scales ranging from the smallest cores to filaments and giant molecular clouds. These insights will help us better understand the processes shaping the chemistry of star and planet formation.  

\section*{Acknowledgements}
We thank Joseph Hora, Gary Melnick and Paola Caselli for insightful discussions, and the anonymous referee for their constructive feedback that helped improve the manuscript clarity. We also thank the SPHEREx space telescope engineering and science teams for promptly making their data public in an easy-to-access format. We thank IPAC/IRSA for hosting the data.

This research has made use of data from the Herschel Gould Belt survey (HGBS) project, a Herschel Key Programme jointly carried out by SPIRE Specialist Astronomy Group 3 (SAG 3), scientists of the PACS Consortium, and scientists of the Herschel Science Center.

\section*{Data Availability}
This analysis is entirely based on public data from SPHEREx, and has used archival data from WISE and Herschel in figures. The simulations use the publicly available codes RADMC-3D and \texttt{OpTool}. 

\bibliography{PASPsample701}{}
\bibliographystyle{aasjournal}

\appendix
\section{Comparing SPHEREx \tauice\ to Literature for L1544 Background Stars}
\label{subsec:bgstars}
\begin{figure*}[h!]
    \centering
    \includegraphics[width=0.36\linewidth]{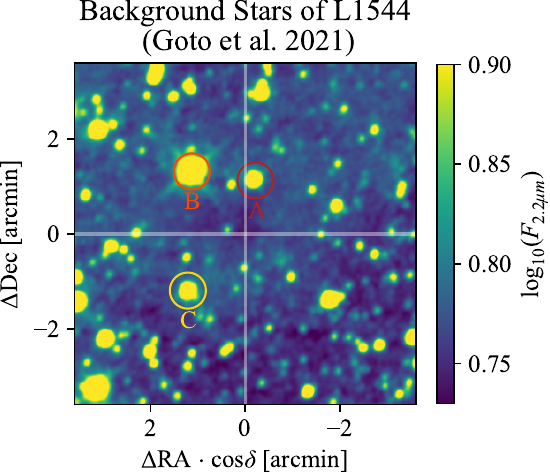} 
    \hspace{15pt}
    \includegraphics[width=0.31\linewidth]{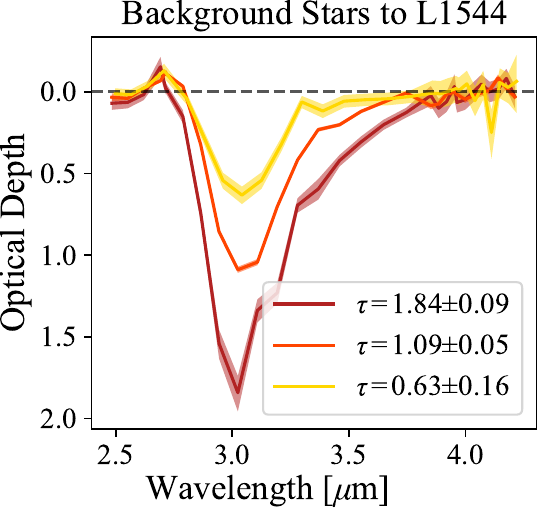}
    \caption{Water ice absorption toward three background stars in L1544 as studied by \cite{Goto2021}. The colour of the circular marker on the WISE image in the left panel corresponds to the colour used to plot the optical depth spectrum on the right panel.}
    \label{fig:l1544-bgstars}
\end{figure*}

\cite{Goto2021} took spectra of five background stars toward L1544 using SpeX/IRTF, and computed \tauice\ for water and methanol for three of the stars. We use the SPHEREx images to perform aperture photometry (in a 2-pixel, 12.4 arcsec radius aperture) for each source to create their spectra and measure \tauice. Note that these are point sources, not extended coreshine, and hence we use full apertures instead of nearest-pixel matching. Results are shown in Fig~\ref{fig:l1544-bgstars}.

Stars marked A, B, C above correspond to stars 1, 3, 5 respectively in \cite{Goto2021}, with reported \tauice\ of $1.99 \pm 0.25$, $1.21 \pm 0.03$ and $0.67 \pm 0.15$ respectively. Our measurements from SPHEREx are shown in Fig~\ref{fig:l1544-bgstars}. These are therefore consistent to within 10\% for the values of \tauice, and fully consistent within errors (and including an extra bandpass correction factor $\approx 1-1.1$, \citealt{Hora2026}). This serves as an independent check on the ability of SPHEREx to probe \tauice\ in dense cores such as L1544 with background star sightlines, and rules out the possibility of any significant instrument-induced inconsistencies between observations and models.

\section{Effect of Diffuse Foreground Flux}
\label{subsec:foreground}
In constructing the ice maps, and more importantly, in inferring the spatial trends, we assume that for all pixels in the cloud, the flux is dominated by the coreshine flux. However, typical sightlines would have background ($bg$) as well as foreground ($fg$) contamination in the surface brightness flux from interstellar dust and gas along the line-of-sight. For a typical core, the background flux will be severely extincted through the central dense regions of the core and can thus be neglected relative to coreshine. However, the foreground still acts as an unchanged additive component.

To estimate the foreground, we need to study an off-core position where micron-sized dust grains and hence coreshine is expected to be absent. We also need to be close to the actual core as the foreground varies on larger spatial scales. Hence for each case, we look at off-core positions $3-6$ arcminutes away that also fall outside the $N_{\rm H_2}=5\times 10^{21}$ cm$^{-2}$ contour (which marks the boundary of coreshine and ice formation for all four cores in Fig~\ref{fig:prestellar-full-1}~and~\ref{fig:prestellar-full-2}). Let the surface brightness at this position be denoted as $I_{\rm off-core}$. We find this to be nearly constant across wavelengths since ice does not form in lower-density diffuse gas to create an absorption feature. Then the foreground flux, $I_{\rm fg}$, must satisfy the two conditions to keep fluxes physical:
\begin{enumerate}
    \item $I_{\rm off-core} = I_{\rm bg}+I_{\rm fg}$, and since the background is non-zero (and likely high along low Galactic latitude sightlines), $I_{\rm fg} < I_{\rm off-core}$ for all off-core positions. We find the minimum typical off-core flux value $I_{\rm off-core} = 0.08$ MJy/Sr, so $I_{\rm fg} < 0.08$.
    
    \item $I_{\rm fg} < I_{\rm core, total\ observed}$ for all core positions and wavelengths. This is effectively saying that for the coreshine flux to remain positive (physical), the foreground flux cannot be higher than the total observed flux for any wavelength.
\end{enumerate}

\begin{figure*}[h!]
    \centering
    \includegraphics[width=0.8\linewidth]{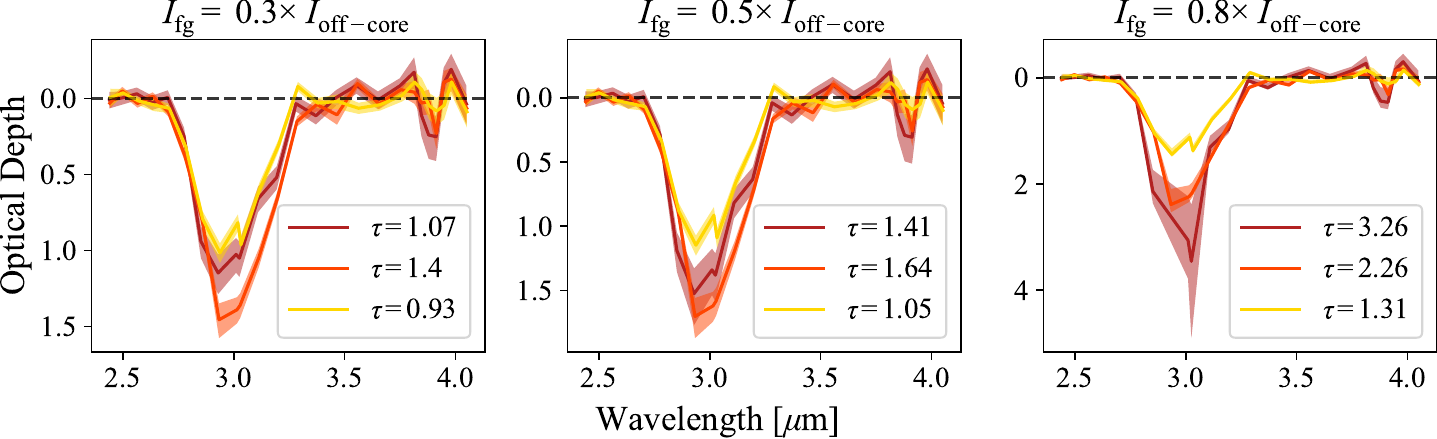} \vspace{10pt}\\
     \includegraphics[width=0.8\linewidth]{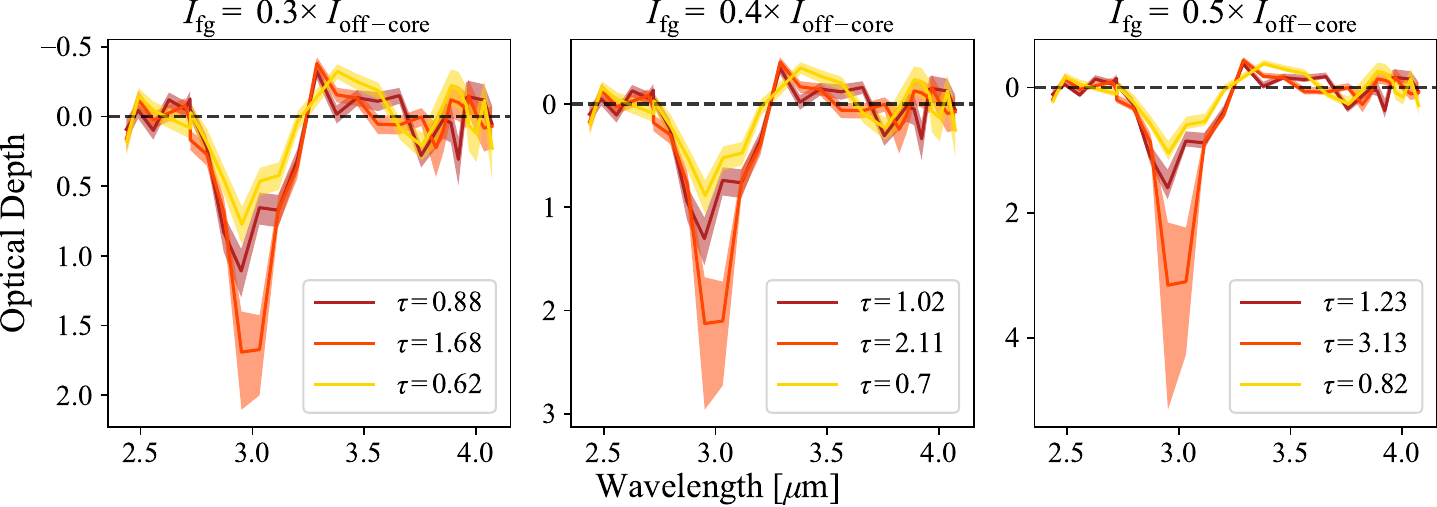}
    \caption{Ice absorption spectra at the three points marked in Fig~\ref{fig:prestellar-full-1} for CrA 151 (top row) and L1544 (bottom row). Brown denotes the spectrum from the cloud centre, orange-red denotes that from an intermediate radius, and yellow is the spectrum from a point in the outskirts. The foreground flux $I_{\rm fg}$ is set to various fractions of the total off-core flux $I_{\rm off-core}$, and subtracted from the observed flux in the cores to compute the new spectra and optical depths. We see for CrA 151 only very high values of $I_{\rm fg}$ can reverse the trend of \tauice\ being lower at centre than at surrounding radii, and for L1544 no assumed $I_{\rm fg}$ can do that. The highest allowed value for the assumed $I_{\rm fg}$ is such that the foreground never exceeds the total observed flux at any position or wavelength}
    \label{fig:foreground-subtraction}
\end{figure*}

For the two dense cores CrA 151 and L1544, we wish to check if the surprising spatial pattern of reduced ice absorption at the centre can be explained by assuming a foreground that needs to be corrected for i.e. subtracted. Hence we estimate the intrinsic flux by subtracting different possible values of $I_{\rm fg}$ from the observed flux and re-calculate \tauice. In each case, $I_{\rm fg}$ is assumed to be some fraction of $I_{\rm off-core}$. The fraction is increased till it hits the second condition. 

The results for the three points in Fig~\ref{fig:prestellar-full-1} is shown in Fig~\ref{fig:foreground-subtraction}. We see that even with a foreground floor subtraction, the spatial trend of \tauice\ being smaller at the centre than at surrounding radii remains the same, except for CrA 151 for high $I_{\rm fg}$. At that point, however, CrA 151 shows absolute values of \tauice$\approx3-4$ incompatible with all our models (\tauice$\lesssim2$) in \S~\ref{subsec:radmc-main-model}, indicating a conservatively overestimated foreground. The trend therefore is most likely physical and not due to an unknown diffuse foreground. We do not include similar figures for the other two cores where \tauice\ peaks at the centre, but check and find that different levels of foreground subtraction do not change the spatial trend.

\section{Physical Setup of the Analytical Model}
\label{appendix:analytical_model}

For the analytical model, we compute the continuum flux and ice-band flux using the respective opacities for absorption ($\kappa_{\mathrm{abs}}$), scattering ($\kappa_{\mathrm{sca}}$), and total extinction ($\kappa_{\mathrm{ext}} = \kappa_{\mathrm{abs}}$+ $\kappa_{\mathrm{sca}}$). 
The opacity values are listed below, and are taken from the \texttt{OpTool} values: 
\begin{itemize}
    \item Continuum (2.1--2.4 $\mu$m): $\kappa_{\rm abs}=2000$ cm$^2$ g$^{-1}$, $\kappa_{\rm sca}=5000$ cm$^2$ g$^{-1}$
    \item H$_2$O Ice (2.9--3.1 $\mu$m): $\kappa_{\rm abs}=6000$ cm$^2$ g$^{-1}$, $\kappa_{\rm sca}=5000$ cm$^2$ g$^{-1}$
\end{itemize}

These are also broadly consistent with published values in literature from Table 6 of \cite{LiDraine2001}, the coreshine-specific albedo range of $\sim0.7-0.9$ \citep{Andersen2013}, and thick ice mantle covered grain opacities from \cite{OH1994}.   
Ice does not add any significant scattering on top of the continuum scattering by dust grains, but considerably increases absorption due to the O---H stretch.

We now compute the mean free path of a photon in a cloud to check how physically useful the single-scattering model is. Assuming a constant gas-to-dust mass ratio of 100, the mean free path can be computed as 

\begin{equation}
    l=\frac{1}{\kappa\rho} \approx 0.08\ \rm{pc}\ \times \left(\frac{\kappa}{1000\ cm^2/g}\right)^{-1} \left(\frac{n_{\rm H_2}}{10^5\ cm^{-3}}\right)^{-1}
\end{equation}
All the cores in our sample and in our models are of radius $\sim0.2$ pc. Due to the steep density drop-off of the Bonnor-Ebert profile, the opacity is dominated by the central flat region, $\tau \approx \kappa\rho_0r_0$. This means that a photon is expected to undergo $\lesssim1$ scattering event through its path in the cloud for $\kappa=2000$ and $n_{\rm H_2}=10^5$ cm$^{-3}$. Hence the single scattering scenario in the model is a reasonable assumption. However, for the case where the central density is significantly high up to $10^7$ cm$^{-3}$, $\tau \approx 30$ and the assumption does not hold.

\begin{figure}[!ht]
\centering
\resizebox{0.5\columnwidth}{!}{%
\begin{circuitikz}
\tikzstyle{every node}=[font=\fontsize{10.8pt}{14.1pt}\selectfont]
\draw [ fill={rgb,255:red,255; green,255; blue,255}, fill opacity=0.54] (4.25,12.375) circle (3.375cm);
\node [font=\fontsize{10.8pt}{14.1pt}\selectfont, fill={rgb,255:red,255; green,255; blue,255}, fill opacity=1, text opacity=1, inner xsep=0.080cm, inner ysep=0.085cm, rounded corners=0.000cm] at (5.875,12.625) {$s$};
\node [font=\fontsize{10.8pt}{14.1pt}\selectfont, fill={rgb,255:red,255; green,255; blue,255}, fill opacity=1, text opacity=1, inner xsep=0.080cm, inner ysep=0.085cm, rounded corners=0.000cm] at (4.875,12.125) {$b$};
\node [font=\fontsize{10.8pt}{14.1pt}\selectfont, fill={rgb,255:red,255; green,255; blue,255}, fill opacity=1, text opacity=1, inner xsep=0.080cm, inner ysep=0.085cm, rounded corners=0.000cm] at (6.25,13.125) {$u$};
\draw [ fill={rgb,255:red,0; green,0; blue,0}, fill opacity=0.46] (5.375,13.25) rectangle (5.625,12.75);
\draw [ fill={rgb,255:red,0; green,0; blue,0}, fill opacity=1] (4.25,12.375) circle (0.03cm);
\draw [{Bar[scale=1.5]}-{Bar[scale=1.5]}, short] (4.3,12.375) -- (5.45,12.375);
\draw [ color={rgb,255:red,8; green,94; blue,230}, draw opacity=1, line width=0.9pt, -{Latex[scale=1.5]}, ] (5.5,16.75) -- (5.5,13);
\draw [ color={rgb,255:red,4; green,95; blue,230}, draw opacity=1, line width=0.9pt, -{Triangle[scale=1.5]}, ] (5.5,12.875) -- (5.5,8.375);
\draw [ color={rgb,255:red,240; green,153; blue,13}, draw opacity=1, line width=0.9pt, short] (8.75,16.75) -- (0.375,16.75);
\draw [{Bar[scale=1.5]}-{Bar[scale=1.5]}, ] (5.75,13.125) -- (5.75,12.375);
\draw [{Bar[scale=1.5]}-{Bar[scale=1.5]}, ] (6.125,14.125) -- (6.125,12.375);
\draw [ fill={rgb,255:red,228; green,22; blue,22}, fill opacity=0.44, rotate around={-90:(5.5, 14.1875)}] (5.375,14.25) rectangle (5.625,14.125);
\begin{scope}
\clip (3.5,7.625) -- (6.5,7.625) -- (7.375,8.25) -- (4.375,8.25) -- cycle;
\foreach \x in {-2.13,-1.97,...,2.13} {
  \draw[rotate around={45:(5.5,7.875)}] ([xshift=\x cm]5.5,5.75) -- ([xshift=\x cm]5.5,10);
}
\end{scope}
\draw [ fill={rgb,255:red,164; green,146; blue,12}, fill opacity=0.44] (3.5,7.625) -- (6.5,7.625) -- (7.375,8.25) -- (4.375,8.25) -- cycle;
\draw [ fill={rgb,255:red,143; green,11; blue,163}, fill opacity=0.54, rotate around={-90:(5.5, 10.9375)}] (5.375,11) rectangle (5.625,10.875);
\node [font=\fontsize{10.8pt}{14.1pt}\selectfont, fill={rgb,255:red,255; green,255; blue,255}, fill opacity=1, text opacity=1, inner xsep=0.080cm, inner ysep=0.085cm, rounded corners=0.020cm] at (7.125,16.375) {Light source};
\node [font=\fontsize{10.8pt}{14.1pt}\selectfont, fill={rgb,255:red,255; green,255; blue,255}, fill opacity=1, text opacity=1, inner xsep=0.080cm, inner ysep=0.085cm, rounded corners=0.020cm] at (7.25,8.5) {Observer};
\end{circuitikz}
}%
\caption{Setup of the analytical model. The background light source is exactly along the line-of-sight. The top blue arrow show the rays of light from the background light source to the infinitesimal element $ds$ (marked in a grey rectangle) at projected radius $b$ and LoS coordinate $s$. The lower blue arrow shows the ray of light from the same element to the observer at the bottom. The small red rectangle shows an infinitesimal extinction element for the incoming light (we integrate over all of these for the total extinction), and the small purple rectangle shows a similar extinction element for the outgoing light.}
\label{fig:model-diagram}
\end{figure}
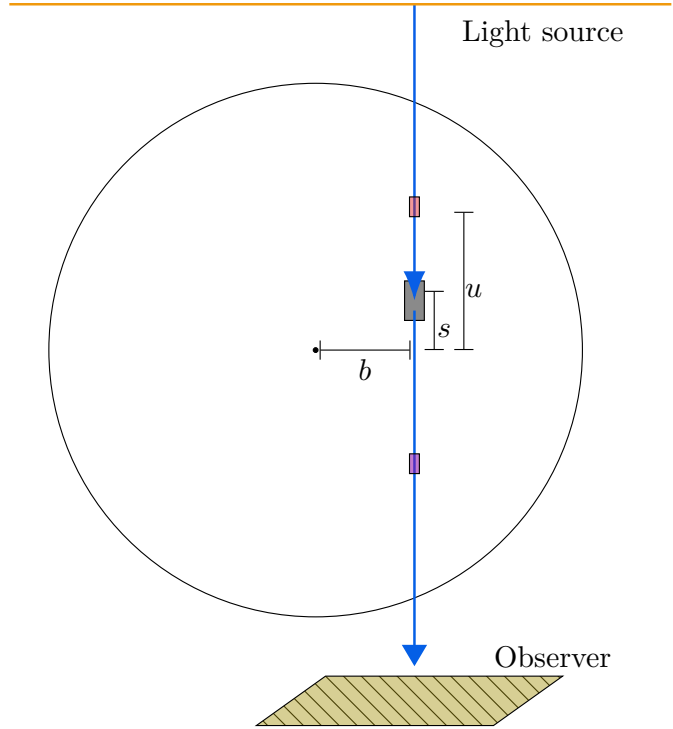

The model we use is shown diagrammatically in Fig~\ref{fig:model-diagram}. We assume a uniform background radiation field with light incident on the core along the line-of-sight (LoS), and the incident light along each line of sight is absorbed and forward-scattered. 

Let $s$ represent the line-of-sight coordinate along which light travels. Let $b$ be the plane-of-sky projected distance from the core centre. For an assumed core radius $R$, $s$ varies between $\pm \sqrt{R^2-b^2}=\pm |s_{\mathrm{max}}|$. We assume light enters the core at $-s_{\mathrm{max}}$ and exits at $+s_{\mathrm{max}}$. Therefore at a given position in the core at $b$ and $s$, the incident light faces a total extinction in the continuum of 
\begin{equation}
    \tau_{\mathrm{in}} = \int_{-s_{\mathrm{max}}}^s \rho\left( \sqrt{b^2+u^2} \right) \kappa_{\mathrm{ext}} du
\end{equation}
If the incident light has some arbitrary intensity $I_{\mathrm{0}}$, the forward scattered light at point $s$ by a column of length $ds$ is given by
\begin{equation}
    \frac{dI_{\mathrm{sca}}}{I_{\mathrm{0}}} = \rho\left( \sqrt{b^2+s^2} \right)\kappa_{\mathrm{sca}} e^{-\tau_{\mathrm{in}}}\cdot ds
\end{equation}
This scattered light faces another extinction column before emerging out of the cloud
\begin{equation}
    \tau_{\mathrm{out}} = \int^{+s_{\mathrm{max}}}_s \rho\left( \sqrt{b^2+u^2} \right) \kappa_{\mathrm{ext}} du
\end{equation}
Thus the final light emitted by the column of length $ds$ is
\begin{equation}
    \frac{dI_{\mathrm{coreshine}}}{dI_{\mathrm{sca}}} = e^{-\tau_{\mathrm{out}}}
\end{equation}
We can integrate to get the total emitted intensity
\begin{equation}
    I_{\mathrm{coreshine}} = \int^{+s_{\mathrm{max}}}_{-s_{\mathrm{max}}} dI_{\mathrm{coreshine}}
\end{equation}

Using the values of $\kappa$ for continuum gives us the continuum intensity $I_{\rm cont}$ due to scattering at each point $b$, allowing us to construct the radial profile. Changing these to $\kappa$ for the ice band gives us the ice-band flux $I_{\mathrm{ice}}$. We can calculate the radial continuum profile and the radial ice absorption profile of 
$$ \tau_{\rm ice} = -\ln \left(\frac{I_{\mathrm{ice}}}{I_{\mathrm{cont}}}\right)$$ as a function of projected radius $b$. This is then explored for various parameter choices in Fig~\ref{fig:model-pre}.

\end{document}